\documentclass[12pt,preprint,revtex4]{emulateapj}
\usepackage{natbib}
\usepackage{graphicx}
\usepackage{pdflscape}
\usepackage{amsmath}
\usepackage{xcolor}
\usepackage{url} 
\usepackage{placeins}
\usepackage{longtable}
\usepackage[caption = false]{subfig}

\shorttitle{CATSUP for FGK-Stars Within 30 pc}
\shortauthors{Hinkel et al.}

\def\lg{{$\log$(g)\,}}

\def\aaa{{\,$\text{\AA}\,$}}

\newcommand{\Chandra}{{\it Chandra}}

\newcommand{\ROSAT}{{\it ROSAT}}

\newcommand{\XMM}{{\it XMM}}

\newcommand{\teff}{$T_{\rm eff}$\,}

\newcommand{\Msun}{\hbox{$\mathcal{M}^{\rm N}_{\odot}$}}

\def\gtaprx{ \mathrel{ \vcenter{
      \offinterlineskip \hbox{$>$}
      \kern 0.3ex \hbox{$\sim$}    } } }

\def\ltaprx{ \mathrel{ \vcenter{
      \offinterlineskip \hbox{$<$}
      \kern 0.3ex \hbox{$\sim$}    } } }
      
\def\aj{{AJ}}
\def\apj{{ApJ}}
\def\apjs{{ApJS}}
\def\apjl{{ApJL}}

\def\aap{{A\&A}}
\def\pasp{{PASP}}

\def\mnras{{MNRAS}}
\def\nat{{Nature}}
\def\pasj{{PASJ}}

\begin{document} 

\title{A Catalog of Stellar Unified Properties (CATSUP) for 951 FGK-Stars Within 30 pc}

\author{
  Natalie R. Hinkel\altaffilmark{1},
  Eric E. Mamajek \altaffilmark{2,3},
  Margaret C. Turnbull\altaffilmark{4},
  Ella Osby \altaffilmark{5},
  Evgenya L. Shkolnik\altaffilmark{5},
  Graeme H. Smith\altaffilmark{6},
  Alexis Klimasewski \altaffilmark{3,},
  Garrett Somers\altaffilmark{1},
  Steven J. Desch \altaffilmark{5}
}
\email{natalie.hinkel@gmail.com}
\altaffiltext{1}{Department of Physics \& Astronomy, Vanderbilt University, Nashville, TN 37235, USA}
\altaffiltext{2}{Jet Propulsion Laboratory, California Institute of Technology, M/S 321-100, 4800 Oak Grove Drive, Pasadena, CA 91109, USA}
\altaffiltext{3}{Department of Physics and Astronomy, University of
  Rochester, Rochester, NY 14627-0171, USA}
\altaffiltext{4}{Global Science Institute, P.O. Box 252, Antigo, WI 54409, USA}
\altaffiltext{5}{School of Earth \& Space Exploration, Arizona State University, Tempe, AZ 85287, USA}
\altaffiltext{6}{University of California Observatories and Department of Astronomy \& Astrophysics, University of California, Santa Cruz CA 95064}

\begin{abstract}

Almost every star in our Galaxy is likely to harbor a terrestrial planet, but accurate measurements of an exoplanet's mass and radius demands accurate knowledge of the properties of its host star. The imminent TESS and CHEOPS missions are slated to discover thousands of new exoplanets. Along with WFIRST, which will directly image nearby planets, these surveys make urgent the need to better characterize stars in the nearby solar neighborhood ($<$ 30 pc). We have compiled the CATalog of Stellar Unified Properties (CATSUP) for 951 stars, including such data as: Gaia astrometry; multiplicity within stellar systems; stellar elemental abundance measurements; standardized spectral types; Ca II H and K stellar activity indices; GALEX NUV and FUV photometry; and X-ray fluxes and luminosities from ROSAT, XMM, and Chandra. We use this data-rich catalog to find correlations, especially between stellar emission indices, colors, and galactic velocity. Additionally, we demonstrate that thick-disk stars in the sample are generally older, have lower activity, and have higher velocities normal to the galactic plane. We anticipate CATSUP will be useful in discerning other trends among stars within the nearby solar neighborhood, for comparing thin-disk vs. thick-disk stars, for comparing stars with and without planets, and for finding correlations between chemical and kinematic properties.

\end{abstract}
\keywords{solar neighborhood --- stars: fundamental parameters --- catalogs}

\section{Introduction}
\label{intro}

The acceleration of exoplanet detections has shifted the field from one of discovery to one of characterization.
Surveys are rapidly becoming complete with respect to Jupiter-sized planets:
occurrence rates of gaseous giant planets with masses $> 50 \, M_{\oplus}$ and periods $< 10$ years is $\sim 14\%$ 
\citep{Mayor11}. 
Even the discovery of Earth- to super Earth-sized (radii $0.5 - 4 \, R_{\oplus}$) planets is becoming routine, with an 
occurrence rate $\sim 1$ per small star inferred \citep[][and references therein]{Dressing13}.
The {\it Kepler} mission has established that planets are nearly ubiquitous around stars, and even multi-planet systems
are common \citep{Batalha13}. 
{\it Kepler} has discovered about 550 rocky exoplanets, including 9 that orbit in their stars' habitable zones
(Kepler press release May 10, 2016).
Scientific progress in the field of exoplanets has advanced so much, it is possible to go beyond asking where the planets are,
to asking what they are made of, a necessary first step to assessing habitability. 
Finding and characterizing exoplanets are the goals of future space mission such as 
the Transiting Exoplanet Survey Satellite (TESS), the CHaracterizing ExOPlanets Satellite (CHEOPS), 
the PLAnetary Transits and Oscillations of stars (PLATO) mission, and the Wide-Field InfraRed Survey Telescope (WFIRST).

It is a truism in the exoplanet community that to know the planet one must know the star. 
Detection of exoplanets has always depended on good characterization of the star. This was true of 51 Pegasi 
\citep{Gray97}, to the recent case of HD 219134h, where a planet inferred by \citet{Vogt15} has been found to have a
period equal to the star's rotation period, making it an artifact of stellar activity \citep{Johnson16}. 
To meaningfully constrain exoplanet compositions, one needs to distinguish a bulk density of $2 \, {\rm g} \, {\rm cm}^{-3}$
\citep[like the densities of Ganymede and Titan, made of rock and ice per][]{Showman99} from one of $5 \, {\rm g} \, {\rm cm}^{-3}$ 
\citep[like the densities of Mercury, Venus, and Earth, with metal cores and rocky mantles per][]{Wanke81}. To measure density at this precision requires constraining 
planetary mass and radius, and therefore stellar mass and radius, to within about 10\% \citep{Unterborn16}.
In order to move beyond simple questions of whether a planet is rock with ice or rock with a large metal core, inclusion of 
elemental ratios, inferred from the host star, must be included. 
The habitability of the planet, and the detectability of life through atmospheric gases, depend on the state of the atmosphere,
which is sensitive to high-energy (X-ray and ultraviolet [UV]) emission from the star.
At every step, characterization of an exoplanet requires comprehensive information about the host star. 

A wealth of information currently exists for the Sun-like (main-sequence, FGK-type) stars within 30 pc that will be the highest 
priorities for exoplanet searches and observations. We examine stars within a radius 30pc to ensure a certain quality sample of stars that is ``complete" for the targets of highest priority direct imaging missions (early/mid-KV and brighter stars, V$>$7-8 at the faintest). Beyond that distance, there is a decreased ability to detect stellar companions, which results in plummeting quality at the cost of exponentially more work in vetting the data from other sources on fainter and more distant stars.
Yet, despite the availability of useful repositories such as Vizier \citep{Ochsenbein00}, PASTEL \citep{Soubiran16}, 
and TOPCAT \citep{Taylor05}, no single database contains all the disparate physical and chemical information needed to
thoroughly characterize nearby Sun-like stars.
What information exists is often spread across several non-standardized databases.

In response, we have created a CATalog of Stellar Unified Properties (CATSUP) where we combine important stellar information with the goal of expediting characterization of planetary systems, their interior structure, and overall habitability. CATSUP incorporates available stellar information relating to astrometry, multiplicity, stellar activity, and stellar abundances with new, unpublished, high-energy emission data. Specifically, we have included new far-UV (FUV) and near-UV (NUV) emission data from Galaxy Evolution Explorer (GALEX). We have also added X-ray data from the ROSAT, XMM, and Chandra missions that have been combined using a new methodology and put on a uniform baseline. Both UV and X-ray information is important when characterizing a planetary system and for understanding exoplanet atmospheres and their possible evaporation. The number of stars and breadth of data within CATSUP allows access to data that was previously difficult or inaccessible. Even when specific stars aren't listed in the database, data can be cross-correlated in order to form a benchmark from which proxies can be determined based on similar, nearby stars. While we have chosen not to focus specifically on planetary properties within CATSUP, the information for solar neighborhood stars can be utilized to better characterize planets and their host stars. We particularly focus on collating data to assess planetary habitability, to help narrow the observational field, and to optimize the search for Earth-like planets, a particularly important goal as telescope time for follow-up observations is limited.

In this paper we discuss the creation and contents of CATSUP, a catalog of 951 nearby ($< 30$ pc) FGK main-sequence stars
containing an array of datasets relevant to exoplanet detection and characterization. For data already available in the literature, we will briefly summarize the details in an effort to maintain a holistic view of the CATSUP database. 
We begin with the high-precision astrometric measurements by the Gaia mission \citep{Gaia16}
as the basis for the catalog (Section \ref{gaia}).
We include currently available data on system multiplicity from ExoCat \citep[][ see Section \ref{exocat}]{Turnbull15}, 
stellar abundance measurements from the Hypatia catalog \citep[][ see Section \ref{hypatia}]{Hinkel14}, collated spectral types (Section \ref{spt}),
and Ca {\sc ii} H and K stellar activity indices \citep[][ Section \ref{stactivity}]{Smith11}.
We present UV emission data from GALEX archives (Section \ref{uv}).
Finally, we present X-ray data from the ROSAT, XMM and Chandra missions (Section \ref{xray}).
As an example of the capabilities of the combined database, we present in Section \ref{application} an exploration of the correlation between Ca {\sc ii} H and K indices with color, galactic velocity, and thin-disk vs. thick-disk membership.

\begin{figure}
\begin{center}
 \centerline{\includegraphics[width=9cm]{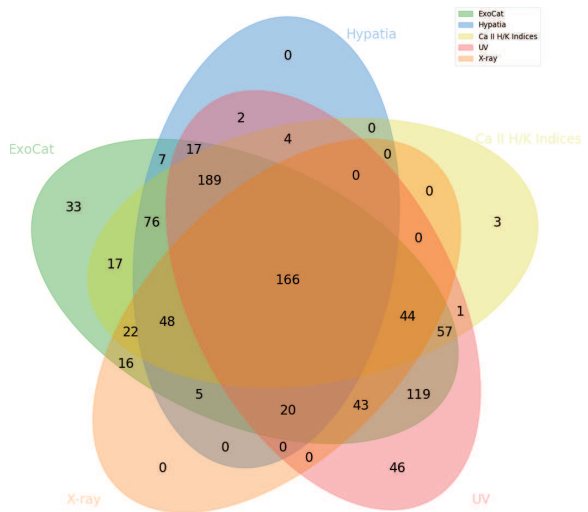}}
  \vspace{-10mm}
\end{center}
  \caption{A Venn Diagram showing the stars that overlap between datasets within CATSUP, such that 166 stars have measurements from all five sources. The total number, or the sum of all the entries, is 951 FGK-type stars within 30 pc. There are 879 stars in ExoCat, 534 in Hypatia, 627 with the Ca II H and K indices, 708 with GALEX UV data, and 364 with X-ray data. Note: All stars are found within Gaia by design. A total of 48 CATSUP stars, or 5\% of the total sample, are currently known to host exoplanets (see Table \ref{params}).}
  \label{venn}
\end{figure}

\section{Gaia}
\label{gaia}
The Tycho-Gaia Astrometric Solution \citep[TGAS, ][]{Gaia16} subset of the Gaia survey provides a convenient and logical starting sample for which data important to both stellar and exoplanet scientific studies can be compiled. Despite the fact that TGAS is $\sim$80\% complete in reference to both {\sc Hipparcos} and {\it Tycho-2} \citep{Arenou17}, the high precision astrometry are valuable for target characterization. Additionally, one of the main reasons for TGAS incompleteness is due to a cut on stars with high parallax errors, namely $>$ 1 mas \citep{Arenou17}, which is acceptable, if not preferable, given our intention. 
Therefore, we started by querying TGAS sources available through Vizier\footnote{\url{http://vizier.u-strasbg.fr/viz-bin/VizieR-3?-source=I/337/tgas}} \citep{Ochsenbein00}. We then placed a 30 pc cutoff, or parallax $>$ 33.33 mas, on the TGAS stars so that we could focus on nearby stars that have both physical and chemical measurements. Note, the distance provided was calculated as the inverse of the parallax, which is a safe assumption for nearby stars as in this case, e.g. \citet{Astraatmadja16}.. Then, using the MK classification of Skiff (see Section \ref{spt}), we included only FGK-type stars in order to maximize completeness.  In this way, CATSUP is able to better achieve catalog completeness by including only nearby main sequence stars with accurate astrometric solutions \citep{Arenou17}. 

The total number of stars with compiled data within CATSUP is 951, shown in Figure \ref{venn}. By summing each of the separate ellipses, you will get the total number of stars in each dataset: 879 stars in ExoCat, 534 in Hypatia, 627 with the Ca II H and K indices, 708 with GALEX UV data, and 364 with X-ray data. Furthermore, we have included a flag within Table \ref{params} to indicate whether a star is known to host an exoplanet at the time of this publication, based on the NASA Exoplanet Archive\footnote{\url{exoplanetarchive.ipac.caltech.edu}. We found that 48 stars within CATSUP are currently known to host exoplanets.}A summary of the parameters available in CATSUP are given in Table \ref{params} -- the full table will be provided via the online journal and through Vizier. In the next sections, we describe the compilation of essential stellar data within CATSUP.

\section{ExoCat}
\label{exocat}

The Nearby Stellar Systems Catalog for Exoplanet Imaging Missions
\citep[a.k.a. ``ExoCat,"][]{Turnbull15} was created for the purpose of supporting the
development of exoplanet direct imaging missions such as WFIRST,
Exo-S, Hab-Ex, and other concepts \citep[][respectively]{Spergel13,Spergel15, Seager15, Mennesson16}.  The
current version of ExoCat contains 2351 entries for Hipparcos stars
within 30pc of the Sun.  ExoCat provides basic observational data
(e.g., Hipparcos astrometry including Equatorial and Galactic
coordinates, parallax, and proper motions, Johnson B and V magnitudes,
and Ks-band magnitudes from 2MASS or converted to 2MASS Ks magnitudes
for bright stars).  Using these data, ExoCat contains derived
estimates of stellar luminosity, effective temperature, stellar radius
and angular size, stellar mass, habitable zone locations and angular
size, photometric fractional planet brightness and V-band magnitude for an
exo-Earth at the Earth-equivalent insolation distance \citep[see][for details]{Turnbull15}.  For bright
stars (V $<$ 7), ExoCat provides separations and delta-magnitudes for
the brightest stellar companion within 10 arcseconds, taken from the
Washington Double Star catalog \citep{Mason01}.

All of these star and system parameters are crucial to understanding
the necessary performance of an exoplanet imaging mission: from
controlling stray light from off-axis companions, to setting
requirements on contrast and speckle stability, inner and outer
working angles, and throughput, to creating a design reference mission
and observing schedule that can be executed within solar avoidance and
other engineering constraints.  ExoCat is under continuous development
in order to provide more accurate system parameters. ExoCat-v1 can be downloaded from the Exoplanets
Exploration Program (ExEP)
website\footnote{\url{http://nexsci.caltech.edu/missions/EXEP/EXEPstarlist.html}}. There are 879 ExoCat stars within CATSUP, where 878 of those stars can also be found in the TESS Input Catalog \citep{Stassun17}. Note the TESS Input Catalog was the only target selection list available out of the three upcoming exoplanet missions (TESS, CHEOPS, and WFIRST).

One especially important and complicated piece of information involves binary and
multiple systems. ExoCat contains a growing set of 
system descriptions identifying components that may not be included in
Hipparcos (or for which the Hipparcos identifier refers to more than
one star); these indicators have been included in CATSUP (see Table \ref{params}).  Each star is noted as either (a) a true single star per a deep literature source (Table \ref{params}, Single = 1), or (b) if known to be a member of a multiple, which component that HIP number is referring to (Table \ref{params}, Component = A, B, etc).  In many cases, the HIP number includes more than one star, either a known or suspected unresolved/very faint companion, which is noted by a ``+" symbol. In a few cases, the ``+" has its own additional entry. If more than one star in the system has its own HIP number, the other associated HIP numbers are listed in the ``HIP2" column. As noted in the original ExoCat paper \citep{Turnbull15}, if the ``Single" column does not equal 1 or if there is a ``+," then it is possible that the system does not correspond to only one star. Namely, it is possible that the HIP entry corresponds to more than one object. If that system comes up as a high priority for a future mission, it should be scrutinized more carefully. Of the stars in our catalog with either known single or binary status, $\sim 45$\% reside in binary or triple systems. This multiplicity fraction matches expectations from volume limited companion surveys, which generally find that 45-50\% of main sequence stars in the mass range 0.7-1.3 M$_{\odot}$ (a close match for the CATSUP catalog) are multiples. As a secondary check, we cross reference CATSUP with the Tycho Double Star Catalog \citep{Fabricius02}, finding that among stars in both catalogs, 49\% have companions. We conclude that the binary fraction of our sample is consistent with expectations.

\section{The Hypatia Catalog 2.1}
\label{hypatia}

\begin{figure*}
\begin{center}
 \centerline{\includegraphics[width=21cm]{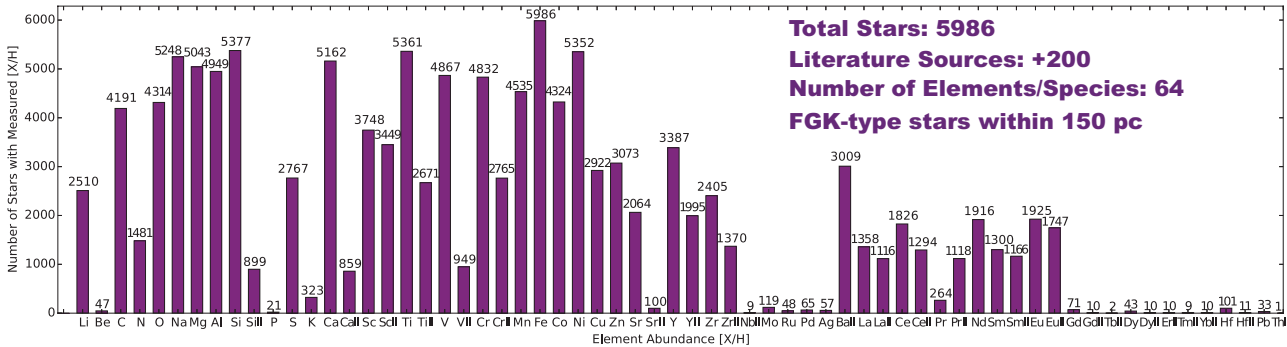}}
\end{center}
  \caption{Number of stars in the Hypatia Catalog that have
    measurements for each element along the x-axis. The total number
    of stars in Hypatia is 5986, which is also the number of iron measurements
    since [Fe/H] is a requirement to be included in Hypatia.}
  \label{hist}
\end{figure*}

The Hypatia Catalog is a composite stellar abundance catalog that is
comprised of multiple literature sources of high resolution spectroscopic data \citep{Hinkel14,
  Hinkel16}. Since the last update published in \citet{Hinkel16}, we
have included 55 additional catalogs from the literature as well as 11 new
elements and species (namely: Si II, Nb II, Pr II, Gd II, Tb II, Dy II, Er II, Tm II, Yb II, Hf II, Th II), as shown in Fig. \ref{hist}. A breakdown of the
added data sets, including information regarding the telescopes,
models, and techniques, is given in Table \ref{update1}-\ref{update4}. There are
currently 64 elements and species measured in 5986 main-sequence stars
within 150 pc of the Sun. Keeping with our naming scheme, we have increased the tenths value of the version number in order to indicate the addition of new data sets. 

In addition to the data sets, we have included a number of new stellar properties within Hypatia and updated the source for some of the existing properties. Namely, we now take advantage of the high-precision RAs, Decs, and parallaxes from Gaia, where applicable, or for $\sim$68\% of the Hypatia Catalog. For the remaining 32\%, we continue our use of \citet{Anderson12} for the RAs, Decs, and parallaxes not available in Gaia. Stellar effective temperature (\teff) and surface gravity (\lg) have been pulled from the PASTEL catalogue \citep{Soubiran16}, in addition to preferentially using their B and V magnitudes where possible. Finally, Hypatia now incorporates the 2MASS identifier. All properties and original sources of reference are listed in Table \ref{params}.

The full Hypatia Catalog Database, including stellar abundances from all individual catalogs, a variety solar normalizations, stellar properties, and planetary properties (where available), can be found at \url{www.hypatiacatalog.com}. Multiple interactive plotting interfaces, in addition to tabular data, can be freely accessed through the website to quickly and easily analyze stellar abundance data, including updated version to standard [X/Fe] vs [Fe/H] plots as a result of new abundance information incorporated into Hypatia. Additionally, data can be downloaded through the terminal for use in personal plotting routines.

For our analysis conducted here, each data set was normalized to the
same solar abundance scale, namely \citet{Lodders:2009p3091}, in order
to minimize systematic differences between the varying
methodologies. During those cases where multiple groups measured the
same element abundance within the same star, the median value was
taken and reported in CATSUP. The range of stellar abundance measurements by different groups is referred to as the spread, a value which often exceeds the individual errorbar \citep{Hinkel14}. Note, unlike previous applications of
the Hypatia Catalog, abundance values were reported even when the
spread in abundance determinations -- or the range of measurements
between groups \citep{Hinkel14} -- was greater than a standardized error. The abundances for 
[Fe/H], [C/H],  [O/H], [Na/H], [Mg/H], [Al/H], [Si/H], [Ca/H], [Ti/H], [V/H], [Cr/H], [Mn/H], [Co/H], and [Ni/H], or elements that have been measured in over 4000 stars in Hypatia, have been included in Table \ref{params}. Additionally, the spread values for all elements in CATSUP can be found in Table \ref{params}, indicated as spXH. Similar to Fig. \ref{hist}, we have included a histogram of the total number of stars in CATSUP for which the 14 elements have been measured in Fig. \ref{cathist}. A total of 534 stars within CATSUP had stellar abundances in Hypatia
2.1. All 534 stars can also be found in the TESS Input Catalog \citep{Stassun17}.

The basic atmospheric parameters, \teff, \lg, and [Fe/H], are very important for the characterization of any stellar sample. Not only do \teff and \lg define the physical conditions of the stellar photosphere, but they are fundamental to stellar abundance measurements. Additionally, stellar \teff directly influences the temperature on a planetary surface while the abundance of [Fe/H] impacts the planet's interior structure and composition. When a CATSUP star was not within Hypatia, the \teff and \lg were sourced from PASTEL; and if it was not found within PASTEL, the \teff and \lg were found within ExoCat. In order to show the parameter space of \teff in CATSUP, we have plotted a Hertzsprung-Russell (HR) diagram of $\log (L/L_{\odot}$) versus \teff, as shown in Fig. \ref{HRdiagram}. We see from this figure that the 270 plotted CATSUP stars all lie along the main sequence, with light scatter possibly due to binaries (see Section \ref{exocat}) or parameter errors. An examination of the spectral types (see Section \ref{spt}) revealed that $<$ 5\% of the CATSUP stars are (sub)giants, meaning that the vast majority of these stars are dwarfs. The data in Fig. \ref{HRdiagram} has been color-coded with respect to the stellar activity indicator, $\log R^{\prime}_{\rm HK}$. While we will discuss this more at length in Section \ref{stactivity}, we see that low-activity stars scatter to brighter luminosities above the main sequence than high-activity stars.

\begin{figure}
\begin{center}
 \centerline{\includegraphics[width=10cm]{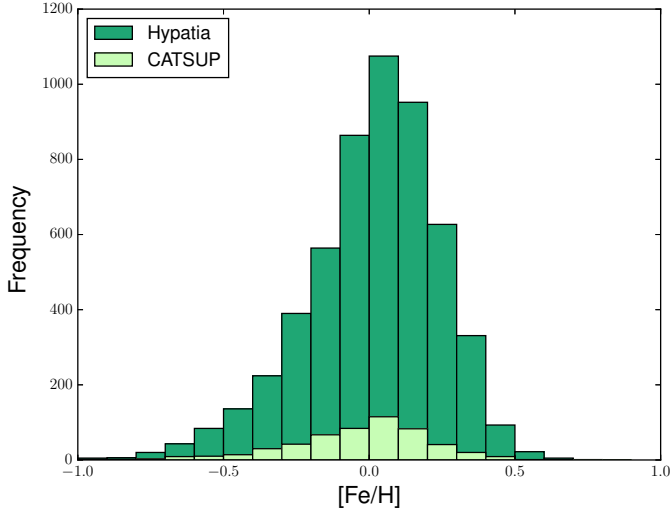}}
\end{center}
  \caption{Frequency distribution showing the number of stars in Hypatia (dark green) and CATSUP (light green) with respect to [Fe/H]. The x-axis is binned in 0.1 dex per the typical error associated with [Fe/H] = $\pm$ 0.05 dex.}
  \label{hypFreq}
    \vspace{3mm}
\end{figure}

\begin{figure}
\begin{center}
 \centerline{\includegraphics[width=10cm]{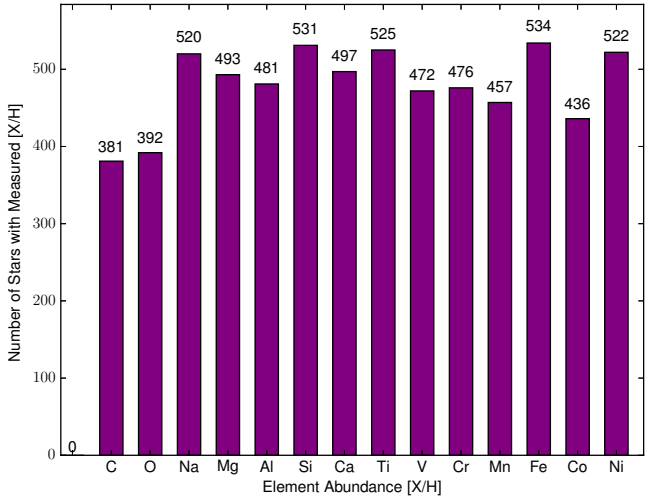}}
\end{center}
  \caption{Similar to Fig. \ref{hist}, here we show the number of stars in CATSUP that have
    measurements for each element along the x-axis. The total number of stars in both CATSUP and Hypatia is 534.}
  \label{cathist}
    \vspace{3mm}
\end{figure}

\begin{figure}
\begin{center}
 \centerline{\includegraphics[width=10cm]{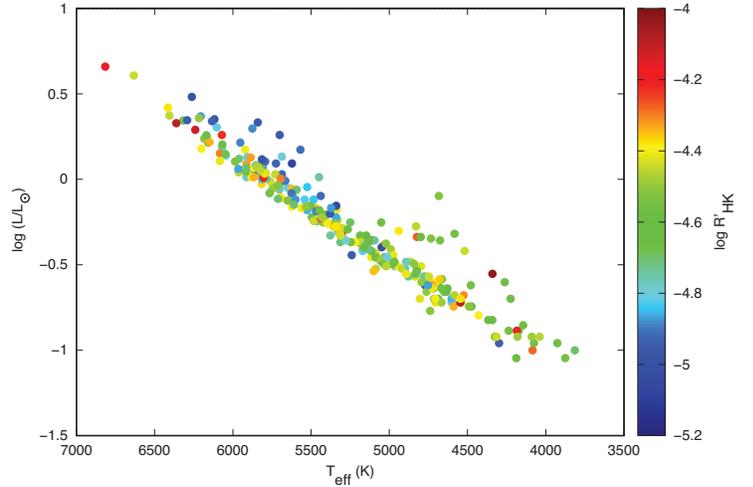}}
\end{center}
  \caption{Hertzsprung-Russell diagram of 270 CATSUP stars, where stars are color coded according to $\log R^{\prime}_{\rm HK}$ emission measurements of stellar activity. This figure was made using Filtergraph \citep{Burger13}.}
  \label{HRdiagram}
    \vspace{3mm}
\end{figure}

We have plotted a distribution of all the Hypatia stars (dark green) with respect to [Fe/H] in Fig. \ref{hypFreq}, where [Fe/H] is given in 0.1 dex bins since the typical associated [Fe/H] error is $\pm$ 0.05 dex. From this plot we see that the majority of stars in Hypatia have solar-like [Fe/H] content. The [Fe/H] distribution of stars within CATSUP is in light green in Fig. \ref{hypFreq}. From the two overlapping samples, we see that the [Fe/H] distribution of the CATSUP subsample mirrors that seen in the full Hypatia Catalog, which are strongly centered around the solar value of iron or [Fe/H] = 0.0 dex. Within both the 30 pc and 150 pc sample of CATSUP and Hypatia, respectively, there is a relatively similar spread in [Fe/H] at all distances which is likely to homogeneous mixture or similar stellar origin within the solar neighborhood.

\section{Spectral Types }\label{spt}

Historically, spectral types have been useful for pooling stars with
similar temperature and luminosity,
identifying spectral anomalies through comparison of spectra to
standard stars, estimating effective temperatures, correcting for the
effects of interstellar reddening on colors, and estimating distances
to stars lacking accurate trigonometric parallaxes. The latter three
reasons are generally not important for studying nearby star samples
as the reddening within the Local Bubble is negligible \citep{Reis11}
and as most of the stars studied here have accurate parallaxes
\citep[e.g.][]{Anderson12, Gaia16}. The modern grid of MK spectral
standard stars is described in \S4.1 of \citet{Pecaut16} and
\citet{Henry02}. We queried Brian Skiff's compendium of MK classifications
\footnote{http://cdsbib.u-strasbg.fr/cgi-bin/cdsbib?2014yCat....1.2023S}
to obtain a fairly complete breakdown of each star's historical spectral classifications and 
notes (e.g. peculiarities, binarity, etc.). 
There were minor shifts in the spectral types of some FGK-standard stars
between the 1940s and 1980s by Philip C. Keenan. In the interest to have the
spectral types as close to being on the modern MK system as possible
(represented by the dwarf standards of \citealt{Keenan89} among
the FGK-type stars), we preferentially adopted spectral types classified
since 1989, and especially those published by expert classifiers using
CCD spectra \citep[e.g.][]{Gray03,Gray06}.

Classifications from the Michigan Spectral Survey
\citep[e.g.][]{Houk75} are only used where necessary, as their
standard star grid varied somewhat from modern grids. This variation results in
systematic offsets sometimes at the $\pm$1.5 subtype level, however these can be corrected following Table 5 of the \citet{Pecaut16} paper. A total of 6 spectral types for the CATSUP stars were adjusted from the Michigan Spectral Survey and have been annotated with an asterisk next to their reference in Table \ref{params}. For some subtypes (e.g. K3V), no adjustment was necessary. Fortunately, most nearby bright stars in our
survey had CCD spectra classified by the NStars project
by \citet{Gray03} and \citet{Gray06}, based on the modern grid of
standards discussed in \S4.1 of \citet{Pecaut16}. The spectral types for the CATSUP stars, and their respective sources, can be found in Table \ref{params}.

\section{Ca II H and K Indices}\label{stactivity}

\begin{figure}
\begin{center}
 \centerline{\includegraphics[width=10cm]{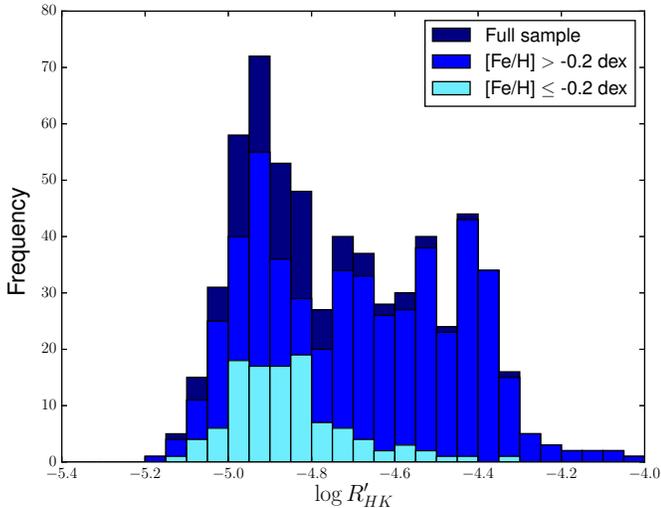}}
\end{center}
  \caption{Frequency distribution of the CATSUP stars with respect to $\log R_{HK}^{\prime}$ for two populations of stars, one with [Fe/H] $>$ -0.2 dex and one with [Fe/H] $<$ -0.2 dex. Both distributions are shown with 0.05 bins of $\log R_{HK}^{\prime}$ in keeping with \citet{Henry96, Gray06, Jenkins11}. } 
  \label{freqRHK}
    \vspace{3mm}
\end{figure}

A widely utilized measure of stellar activity derivable from ground-based
observations has been the amount of chromospheric emission in the cores of the \ion{Ca}{2} 
H and K lines, at 3968 \AA\, and 3933 \AA, respectively. One of the most physically meaningful parameterizations of this emission is
the $R^{\prime}_{\rm HK}$ index that is defined by \citet{Noyes84} as being the chromospheric
flux in the combined H and K lines radiated by a star relative to the bolometric
flux of the star. It is an index that has been corrected for photospheric contributions to the
flux across the wavelengths of the H$_2$ and K$_2$ lines, which are the chromospheric emission (reversals) of the Ca II H and K lines. Early discussions of the behavior
of this index, or similar ones not corrected for a photospheric component, among nearby 
late-type dwarf stars can be found in \citet{Middelkoop82, Hartmann84, 
Noyes84, Soderblom85}, all of whom built on the work of \citet{Wilson78,Vaughan80}.

Values of $\log R^{\prime}_{\rm HK}$ emission measurements, are listed in Table \ref{params}, for 627 stars within CATSUP, Table \ref{params}. These values were selected from a larger data base 
that were compiled from papers in the literature prior to 2011 per \citep{Smith11}, in addition to two stars from \citet{daSilva14}.
The characteristics of this
pre-2011 data base of $\log R^{\prime}_{\rm HK}$ values have been discussed by \citet{Smith10} and \citet{Smith11}, who also describe the manner in which the data base was compiled.
By far the largest literature sources used in the compilation of \citet{Smith11} are the 
following: \citet{Gray03, Gray06, Henry96, Jenkins06, Jenkins08, 
Noyes84, Soderblom85, Soderblom91, Wright04}. 
Where multiple literature values were available they were averaged with equal weights, after 
first applying systematic offsets to values from \citet{Gray03, Gray06}, in order to yield
a homogenized data set. More details are discussed in \citet{Smith11}, whose paper can also
be resorted to for a discussion of possible time variability in the H and K emission, as well as 
references to a number of other smaller literature sources that were used.  All of those stars with Ca II H and K indices can also be found in the TESS Input Catalog \citep{Stassun17}.

An HR-diagram for the CATSUP stars for which Ca II H and K
emission has been measured is shown in Fig. \ref{HRdiagram}, where the datapoints have been
coded on a continuous color-scale according to the value of the
$\log R_{HK}^{\prime}$ emission parameter. This figure illustrates the
main-sequence nature of the majority of the CATSUP stars. There are some
more-evolved stars that scatter by up to $\sim 0.5$ dex in $\log$ (L/L$_{\odot}$)
above the main sequence. Among solar-type dwarfs with $T_{\rm eff} > 5400$ K
these more-evolved stars tend to exhibit relatively low levels of Ca II H and K
emission $\log R_{HK}^{\prime} < -4.8$. Relatively few dwarfs cooler than
5400 K in Fig. \ref{HRdiagram} are found to have such low levels of activity. Consequently,
the values of $-4.8 < \log R_{HK}^{\prime} < -4.5$ that do dominate among the
more-evolved CATSUP stars in the solar-like temperature regime can be considered
closer to the lower levels of activity encountered in the cooler CATSUP stars.
\citet{Wright04} found that many so-called ``Maunder Minimum'' stars
of near-solar temperature but of notably lower activity levels than the
average Sun, are in fact slightly evolved from the main sequence. The
distribution of CATSUP stars with $\log R_{HK}^{\prime} < -4.9$ seems
consistent with the findings in \citet{Wright04}.

The distribution of chromospheric activity within the CATSUP sample is shown
in Fig. \ref{freqRHK}. The pioneering work of \citet{Vaughan80} indicated a bimodal
distribution among nearby dwarf stars, with high and low activity groups being
separated by the so-called Vaughan-Preston gap. This gap can also be seen in
samples discussed by \citet{Middelkoop82, Noyes84, Soderblom85}. 
A bimodality was evinced in the $R_{HK}^{\prime}$ surveys of \citet{Henry96} and \citet{Gray06}. 
The survey of southern hemisphere stars
by \citet{Jenkins08} found that the gap corresponded to a relatively low
percentage of stars with activity levels of $\log R_{HK}^{\prime} \sim -4.7$,
while active and inactive stars had mean levels of $-4.5$ and $-5.0$
respectively. While a corresponding low-activity peak at $\log R_{HK}^{\prime}
\sim -4.9$ is seen for the CATSUP distribution in Fig. \ref{freqRHK}, a possible high
activity peak near $\sim -4.4$ is quite muted. Within the CATSUP sample the
Vaughan-Preston gap does seem to be relatively well populated, with a broad
continuous distribution extending across the activity range from $-4.8$ to
$-4.3$.

\citet{Gray06} found that the distribution of $\log R_{HK}^{\prime}$
differs between stars of different metallicities, with a bimodality largely
being confined to dwarfs with [M/H] $> -0.2$ dex. Within Fig. \ref{freqRHK} we have
similarly divided the CATSUP sample into metallicity groupings of [Fe/H] less
than or greater than $-0.2$ dex. The more metal-rich bin within CATSUP is not as
strikingly bimodal as that shown in Fig. 5 of \citet{Gray06}, while the
metal-poorer distribution in our Fig. \ref{freqRHK} is similar to that found in their
Fig. 5. Stars above the Vaughan-Preston gap are thus largely of metallicity
similar to the Sun, while dwarfs with [Fe/H] $< -0.2$ dex constitute a
significant component of the low-activity peak.

\section{UV Photometry with GALEX}
\label{uv}

An additional stellar activity measure critical to the understanding of exoplanets is the UV light from host stars. As the high energy UV photons alter the planets' atmospheric evolution and photochemistry, interpretations of exoplanetary spectra are affected for both close-in giant planets and habitable zone earths \citep{migu14a, rugh15, luge15, Arney16}. We cross-correlated the CATSUP target catalog with the archived 
UV photometry of the GALEX space mission from the General Release\footnote{The GALEX GR6/7 is available at
http://mastweb.stsci.edu/gcasjobs/} GR6/7. The GALEX satellite imaged
roughly two-thirds of the sky from 2003 to 2013 in two UV bands: the
NUV (1750-2750 \AA) and FUV (1350-1750 \AA).  The
full description of the instrument and its mission is provided by
\cite{morr05} and of the pipeline by \cite{morr07}. The failure of the
FUV detector in 2009 resulted in only NUV imaging for all observations
after this date.  We queried the GALEX archive for all stars in the
catalog using a search radius of 30$\arcsec$. 

The NUV detector response becomes non-linear beyond 34 counts per
second, which occurred for 102 (11\%) of the CATSUP targets. In
the case of non-linearity for these NUV measurements, we report the
pipeline's measured flux density as a lower limit, and
\citet{Shkolnik13} was used as a reference for upper limits for bright
targets in GALEX. In the FUV, the detector's response becomes
non-linear beyond 108 counts per second, but only affected a tiny fraction of the
CATSUP stars. GALEX did not detect 309 (32\%) of the stars in the
FUV for which we report estimated upper limits. Flux densities,
including their upper limits in the case of non-detection and
lower-limits in the case of non-linear detector response are listed in
Table \ref{params}. If a star's NUV or FUV emission was detected by
GALEX, but it was unable to be measured completely due to limited field of
view, the data were deemed unreliable and excluded from the final
catalog. There are 708 stars with UV measurements within CATSUP, where 676 of those stars can also be found in the TESS Input Catalog \citep{Stassun17}. 

\begin{figure*}[t]
\begin{center}
\begin{tabular}{cc}
    \includegraphics[width=3.5in]{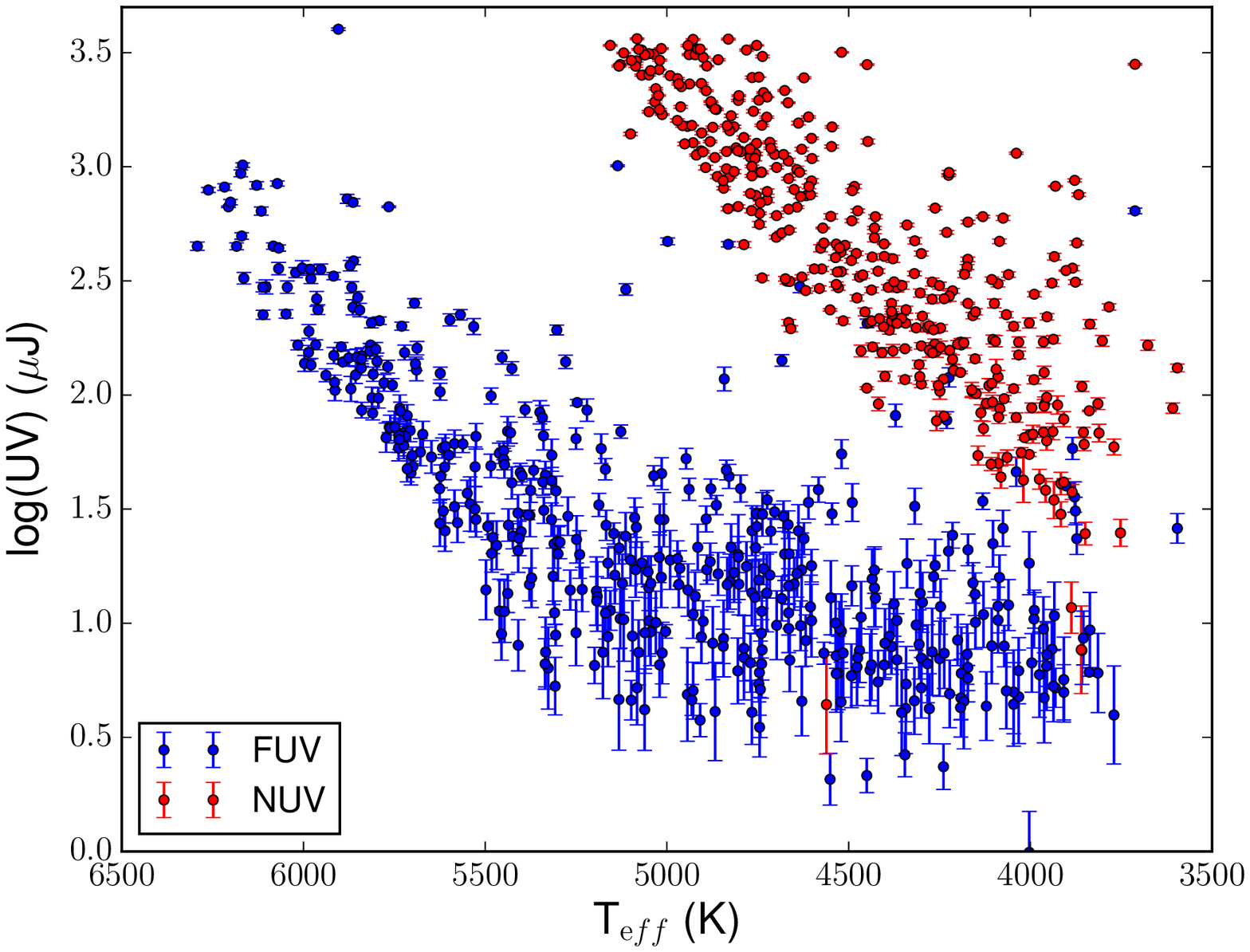}
    \includegraphics[width=3.5in]{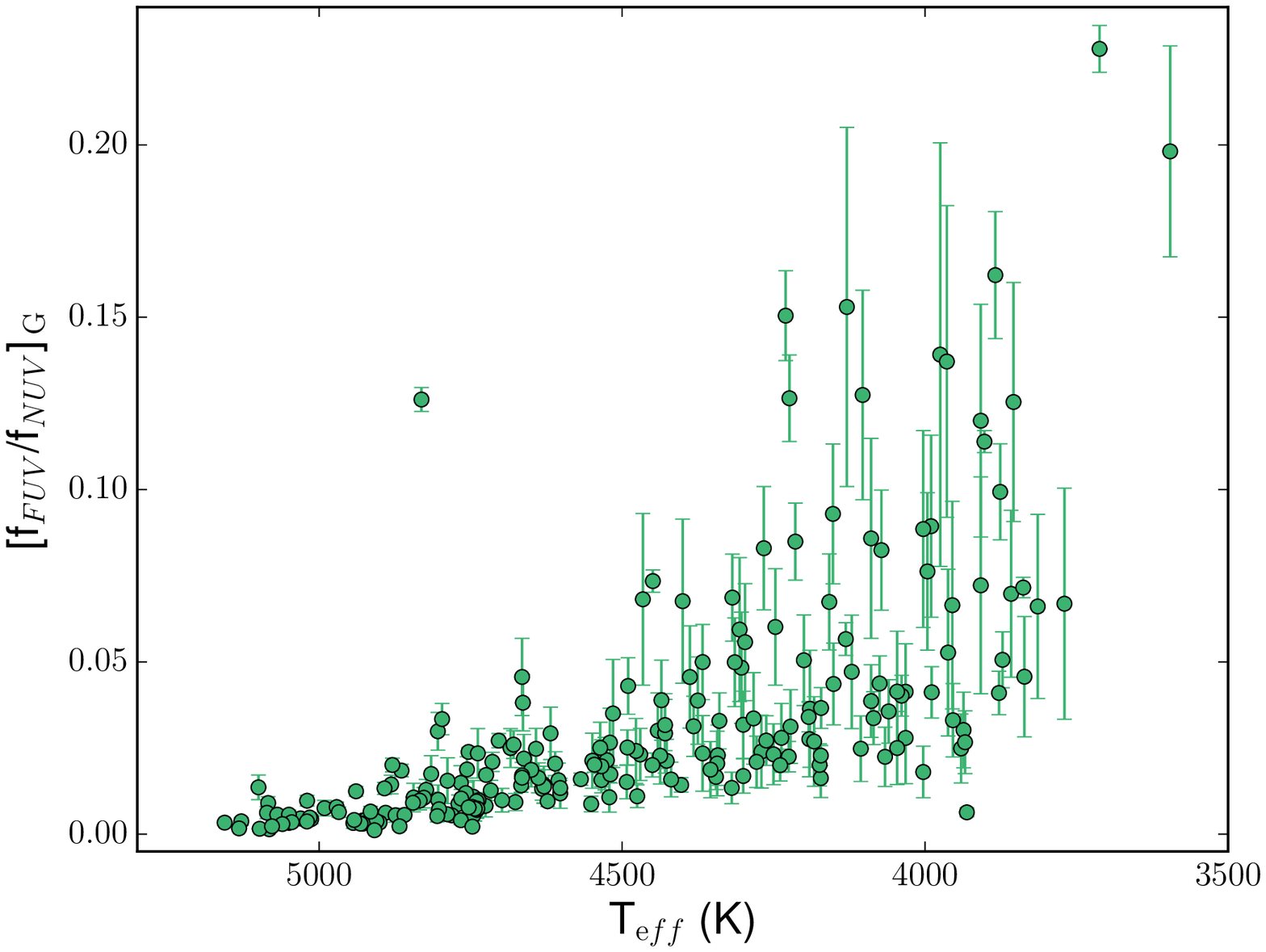}
\end{tabular}
    \caption{Left: The log of the FUV (blue) and NUV (red) GALEX fluxes with respect to effective temperature, \teff. The NUV data are cut off at \teff $<$ 5200 K due to non-linearity in the detector for bright stars. Right: The ratio of the flux densities in the FUV and NUV GALEX band passes with respect to effective temperature, \teff.} 
    \label{fuvnuv}
    \vspace{7mm}
\end{center}
\end{figure*}

In Fig. \ref{fuvnuv}, we show a plot of the effective temperature (\teff) with respect to the logged fluxes for both the FUV (blue) and NUV (red) with individual error bars. When \teff is below 5500 K, the FUV scatters due to the interference with the star's corona, namely that coronal interference is higher than the blackbody curve. Above \teff = 5500 K, the FUV flux is dominated by the photosphere, which is temperature dependent. As the photosphere becomes a smaller fraction of the total FUV bandpass, the flux is increasingly from the chromosphere, the transition region, and the corona, which is dominated by stellar activity and not effective temperature. The NUV is only well detected when \teff $<$ 5200 K, hence the sharp cut-off for the red data points. This temperature cutoff also affects the number of data points in the [f$_{FUV}$/f$_{NUV}$]$_G$ relation, or the ratio of the flux densities in FUV and NUV band passes in GALEX, shown on the right in Fig. \ref{fuvnuv}. The trend towards higher ratios with smaller effective temperature is due to the rapid reduction in NUV photons from the stellar photosphere. The large range in the UV ratio at a given temperature is due to variations in stellar activity.  Per Fig. \ref{fuvnuv}, we note that $\sim$ 10 stars have \teff $<$ 3900 K, the typical cutoff between K and M dwarfs. We have confirmed that their effective temperatures are accurate, such that in nearly all cases, these stars were measured by multiple sources that gave similar temperatures. Since there are no obvious reasons for excluding these stars, we have opted to keep them within CATSUP.

\begin{figure}
\begin{center}
 \centerline{\includegraphics[width=10cm]{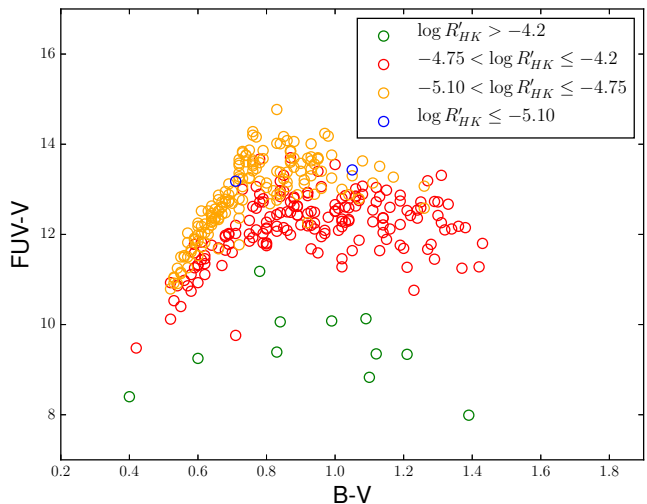}}
\end{center}
  \caption{ A two-color, B-V, diagram for CATSUP stars for which GALEX data
are available. FUV-V is a hybrid color combining the GALEX FUV magnitude and
Johnson V magnitude. Data points are color-coded according to the chromospheric activity
categories defined by \citet{Henry96}. There are few stars in the very
inactive category, but the other three activity categories are well represented
and have clear differences in the FUV-V color at a given B-V.}
  \label{FUV-BV}
  \vspace{3mm}
\end{figure}

  \citet{Smith10} and \citet{Findeisen11} showed the GALEX FUV
magnitude of FGK dwarfs was sensitive to the level of stellar activity as
judged from the strength of the Ca II H and K emission lines. The CATSUP stars
also verify this result. Figure \ref{FUV-BV} presents a two-color B-V diagram for CATSUP stars
for which GALEX measurements of FUV magnitude have been made. It shows a
hybrid color denoted FUV-V, which is obtained by combining the GALEX FUV
magnitude and Johnson V magnitude (following the precepts of \citealt{Findeisen11}),
plotted against Johnson B-V color. To bring out the variation of FUV brightness
with stellar activity, the data points have color-coded according to the
categories defined by \citet{Henry96}, namely,
$\log R_{HK}^{\prime} \le -5.1$ (very inactive),
$-5.1 < \log R_{HK}^{\prime} \le -4.75$ (inactive),
$-4.75 < \log R_{HK}^{\prime} \le -4.2$ (active),
$\log R_{HK}^{\prime} > -4.2$ (very active).
Only a few CATSUP stars fall in the very inactive category, but the other
three activity categories are well represented (Fig. \ref{freqRHK}). Specifically, of the CATSUP stars that had $\log R_{HK}^{\prime}$ measurements, 3\% are categorized as very active, 47\% are active, 49\% are inactive, and 1\% are very inactive. 
These three categories
occupy distinct regions of Fig. \ref{FUV-BV}, and there are clear differences in the FUV-V
color among them at a given B-V. As the degree of chromospheric activity
increases, the values of the FUV-V color decreases due to increasing flux in the
GALEX FUV band arising from stellar active regions. An interested reader is
referred to the papers by \citet{Smith10} and \citet{Findeisen11} for much more detail about this trend.

\section{X-ray data}\label{xray}

X-ray fluxes, luminosities, and fractional luminosities were based on
observations with the {\it Chandra}, {\it XMM}, and \ROSAT\, missions.
The derived X-ray fluxes were converted to a common energy band
(\ROSAT) to facilitate intercomparison of the various X-ray indices.
Additionally, \ROSAT was used as a baseline since much previous work explored rotation-activity relations and age dating using X-ray luminosities form that mission \citep{Mamajek08, Wright11}.

\subsection{Chandra}

The {\it Chandra X-ray Observatory} provides X-ray imaging in peak
energy range $\sim$0.5-7.0 keV, with sub-arcsecond point spread
function, over a $\sim$60-250 arcmin$^2$ field of view
\citep{Weisskopf00}.  We initially queried the CATSUP catalog (J2000
positions from SIMBAD) against the \Chandra\, Source Catalog, Release
1.1 \citep{Evans10,Evans12}, which catalogued X-ray sources in imagery from
the Advanced CCD Imaging Spectrometer \citep[ACIS;][]{Garmire03}.
Chandra positions have astrometric accuracy better than
$\sim$1\arcsec\footnote{http://cxc.harvard.edu/cal/ASPECT/celmon/},
and broadband fluxes (``{\it b}'') in the 0.5-7.0 keV were recorded.
Vetting of the \Chandra\, X-ray counterparts against
the optical astrometry yielded 11 reliable matches. 

\subsection{XMM}

X-ray Multi-Mirror Mission (XMM-Newton) European Photon Imaging Camera
(EPIC) \citep{Struder01} has field of view 30\arcmin\, and covers the
energy range 0.15-15 keV with moderate angular resolution
($\sim$6\arcsec). A query of the CATSUP database (J2000 SIMBAD
positions) with the 3XMM-DR5 catalog of serendipitously detected X-ray
sources \citep{Rosen16} yielded 54 X-ray sources within 90\arcsec. To ease comparison with the
\ROSAT\, X-ray fluxes, the XMM fluxes in bands 1, 2, and 3 were added,
covering 0.2-2.0 keV.

\subsection{ROSAT}

\ROSAT\, (ROentgen SATellite) conducted the \ROSAT\, {\it All-Sky
  Survey} \citep[RASS][]{Voges99, Voges00} during a half-year period
shortly after launch in 1990. The RASS mapped 99.7\% of the sky with
exposure times over 50 seconds in the 0.1-2.4 keV band with the
Position Sensitive Proportional Counter (PSPC), and catalogued 18,811
sources down to an approximate limiting count-rate of 0.05
cts\,s$^{-1}$ \citep{Voges99} in the Bright Source Catalog (RASS-BSC;
1RXS). The faint star extension of the RASS was published as the Faint
Source Catalog (RASS-FSC) of 105,924 X-ray sources \citep{Voges00}. A
reanalysis of the RASS was recently completed by \citet{Boller16},
which resulted in a catalog of 135,000 X-ray sources in the 2nd {\it
  ROSAT} All-Sky Survey source catalogue (2RXS). Until the future
eROSITA mission completes its survey, the 2RXS provides the
astronomical community with the deepest all-sky X-ray survey.  We
adopted the count-rates and hardness ratio (HR1) from \citet{Boller16}
to calculate the energy conversion factor and resultant X-ray flux in
the 0.1-2.4 keV band using the linear trend from \citet{Fleming95}:

\begin{equation}
ECF\,=\,(8.31\,+\,5.30\,{\rm HR1})\,\times\,10^{-12}\,{\rm erg\,cm^{-2}\,ct^{-1}}
\end{equation}
 
Both the original RASS analyses and the revised RASS catalog produced
by Boller et al. have similar positional uncertainties of typically
$\sim$13\arcsec \citep{Boller16}. Experience has shown that an optimal
search radius for matching optical stars with their RASS X-ray
counterparts is 40\arcsec\, \citep{Neuhauser95}. A larger
search radius of 90\arcsec\, was employed in a few cases to retrieve
matches for some of the nearest stars which had large proper motions.

The Second ROSAT PSPC Catalog \citep{ROSAT00} of X-ray sources
detected in {\it pointed} PSPC observations was also queried with both
a 40\arcsec\, and 90\arcsec\, search radius around the positions of
the CATSUP stars. A total of 109 stars had X-ray sources both in the
\ROSAT\, All-Sky Survey and the Second \ROSAT\, PSPC Catalog. As the
pointed observations in the latter catalog had longer exposure times
than the All-Sky Survey, we adopted the PSPC count rates and hardness
ratios HR1 from the latter, and calculated soft X-ray fluxes using the
previously mentioned formula from \citep{Fleming95}.\\

\subsection{X-ray Flux Conversion}

We would like the X-ray fluxes to be on a common system so that
stellar activity levels can be usefully compared between stars.
Comparison of the \XMM, \Chandra, and \ROSAT\, fluxes amongst sources
with detections in multiple surveys produced plots with large scatter
(likely due to X-ray variability) and it was not clear that empirical
flux conversions could be accurately estimated. We decided to use the
WebPIMMS tool, or Portable, Interactive Multi-Mission
Simulator\footnote{heasarc.gsfc.nasa.gov/cgi-bin/Tools/w3pimms/w3pimms.pl}
\citep{Mukai93}, to intercompare the X-ray fluxes between the
observatories. For calculating X-ray flux conversions, we required an
estimate of the \ion{H}{1} column density. The median distance to the
stars in the CATSUP catalog ($d$ $<$ 30 pc) is $\sim$24 pc. Trends of
hydrogen column density versus distance within the Local Bubble are
consistent with neutral hydrogen densities of $n_H$ = 0.1 cm$^{-3}$
\citep{Linsky00}. Hence for a mean distance of 24 pc for our sample
stars, we adopt log\,N(\ion{H}{1}) = 18.8 as a representative hydrogen
column density for the CATSUP sample for the webPIMMS tool.

\begin{figure}
\begin{center}
 \centerline{\includegraphics[width=10cm]{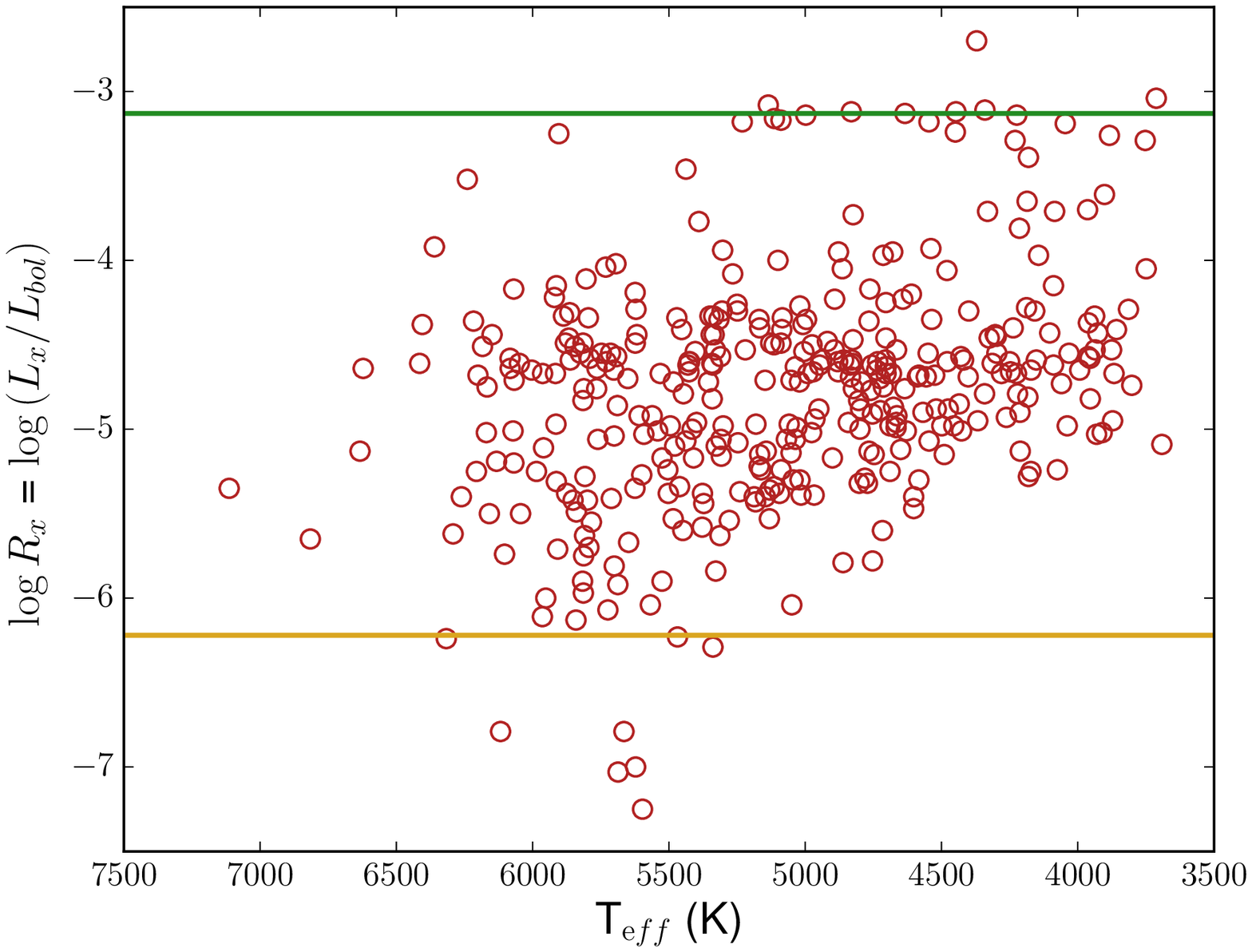}}
\end{center}
  \caption{The fractional X-ray luminosity} ($\log R_x$) with respect to effective temperature (\teff) for stars within CATSUP. The yellow horizontal line is the solar coronal activity value and the green line is X-ray saturation per \citet{Wright11}. 
  \label{xrayteff}
    \vspace{3mm}
\end{figure}

The conversion between X-ray fluxes for one instrument and another
depends on the temperature of the plasma, its chemical composition,
and the intervening hydrogen column.  \citet{Johnstone15} demonstrated
a strong correlation between coronal X-ray temperature and X-ray
surface flux for main sequence stars with masses between $\sim$ 0.2
and $\sim$1.1 \Msun: $\overline{T}_{\rm cor}$ =
0.11\,$F_{X}^{0.26}$. The correlation spanned inactive stars such as
the Sun at solar minimum ($T_X$ $\simeq$ 1.0 MK, $F_X$ $\simeq$
10$^{3.65}$ erg\,s$^{-1}$\,cm$^{-2}$) to very active stars like 47 Cas
B ($T_X$ $\simeq$ 11 MK, $F_X$ $\simeq$ 10$^{7.61}$
erg\,s$^{-1}$\,cm$^{-2}$).  Indeed coronal temperatures scale much
more closely with X-ray surface flux than with X-ray luminosity
($L_X$) or X-ray to bolometric luminosity ratio ($R_X$ = $L_X$/$L_{\rm
  bol}$).  To produce self-consistent X-ray flux conversions between
the \XMM, \Chandra\, and \ROSAT\, data, we iteratively solve for a
consistent combination of $F_X$ and $T_{\rm cor}$ using the
\citet{Johnstone15} power law.  An initial estimate of $T_{\rm cor}$
is adopted (1 MK), an initial conversion from \Chandra\, or \XMM\,
X-ray flux to \ROSAT\, flux is calculated using PIMMS at this
temperature (adopting log\,N(\ion{H}{1}) = 18.8 and solar
composition). Then, an initial X-ray surface flux ($F_X$) is calculated, and
a revised coronal temperature $T_{\rm cor}$ is recalculated.  This
cycle is iterated until $T_{\rm cor}$ changes by less than 0.01 dex
between iterations. The iterative method encountered a few troublesome cases for low
$T_{cor}$ that would not converge, all with $T_{cor}$ $<$ 1 MK. Based
on the observed X-ray fluxes at solar minimum and that observed for
coronal holes, we simply adopted $T_{cor}$ = 1 MK for these cases. 

The final converted fluxes, flux errors, luminosity, and activity from the \XMM, \Chandra, and \ROSAT\, X-ray missions can be found in Table \ref{params}. A total of 364 stars have X-ray measurements within CATSUP, where 363 of those stars can also be found in the TESS Input Catalog \citep{Stassun17}. A plot of the fractional X-ray luminosity $\log R_x$ = $\log \,( L_x/L_{bol} )$, using the median value of $\log R_x$ from the three missions, with respect to \teff is show in Fig. \ref{xrayteff}. The yellow line is the coronal activity value of the Sun while the green line is the empirical X-ray saturation limit \citep{Wright11}. The two lines show that the majority of the CATSUP stars (with X-ray measurements) fall within these two values and that the upper envelope is relatively constant across spectral types, which implies saturation. We also see that the cooler, lower mass stars are more active. The spread in the $\log R_x$ values is likely the result of different rotation rates, since it is expected that equal mass stars would have approximately the same bolometric luminosity. 

\begin{figure}
\begin{center}
 \centerline{\includegraphics[width=10cm]{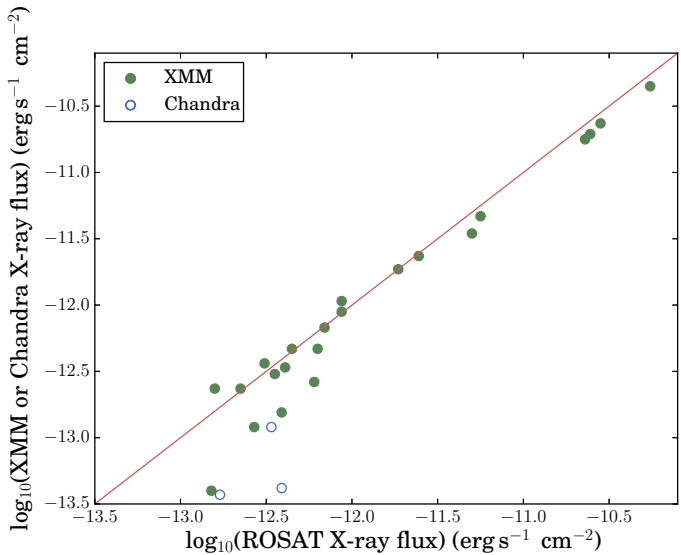}}
\end{center}
  \caption{The X-ray flux from ROSAT (on the x-axis) with respect to X-ray flux measurements {\textit in the same star} from XMM (green, closed circle) or Chandra (blue, open circle). A one-to-one line in overlaid in red.}
  \label{xraycomp}
    \vspace{3mm}
\end{figure}

As a consistency check, we searched Table \ref{params} for stars with X-ray measurements from two different instruments, in order to compare their normalized fluxes. Among our sample, we find 22 stars with both \ROSAT\ and \XMM\ data, and three stars with both \ROSAT\ and \Chandra\ measurements. The derived flux values are plotted relative to one another in Fig. \ref{xraycomp}. For all points, the abscissa is the normalized flux from \ROSAT, and the ordinate can be either flux from \XMM\ (filled green) or \Chandra\ (empty blue). The flux values generally show good agreement for the X-ray-brighter stars, suggesting that our normalization process is reliable when the fluxes are well measured. At the dimmer end, some stars deviate from the one-to-one line, a likely consequence of poorer counting statistics in the shallower \ROSAT\ survey. However intrinsic fluctuations in the X-ray brightness, for example due to solar-type cycles, could also contribute.

\section{Application}
\label{application}

\begin{figure}
    \includegraphics[width=3.5in]{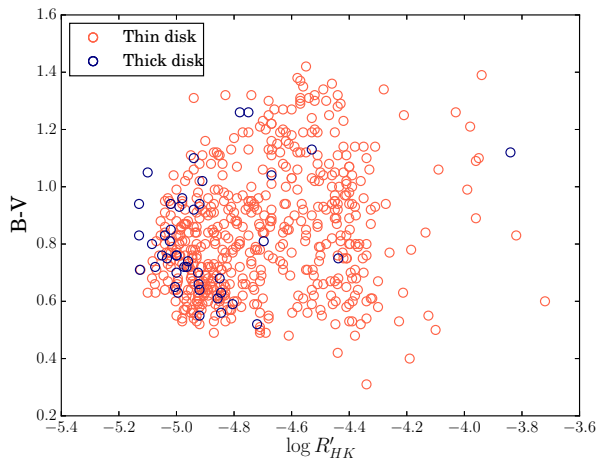}
    \caption{The \ion{Ca}{2} H and K index, $\log R^{\prime}_{\rm HK}$, with respect to B-V color.} 
    \label{activity}
\end{figure}

\begin{figure*}
\center
\subfloat{\includegraphics[width = 3in]{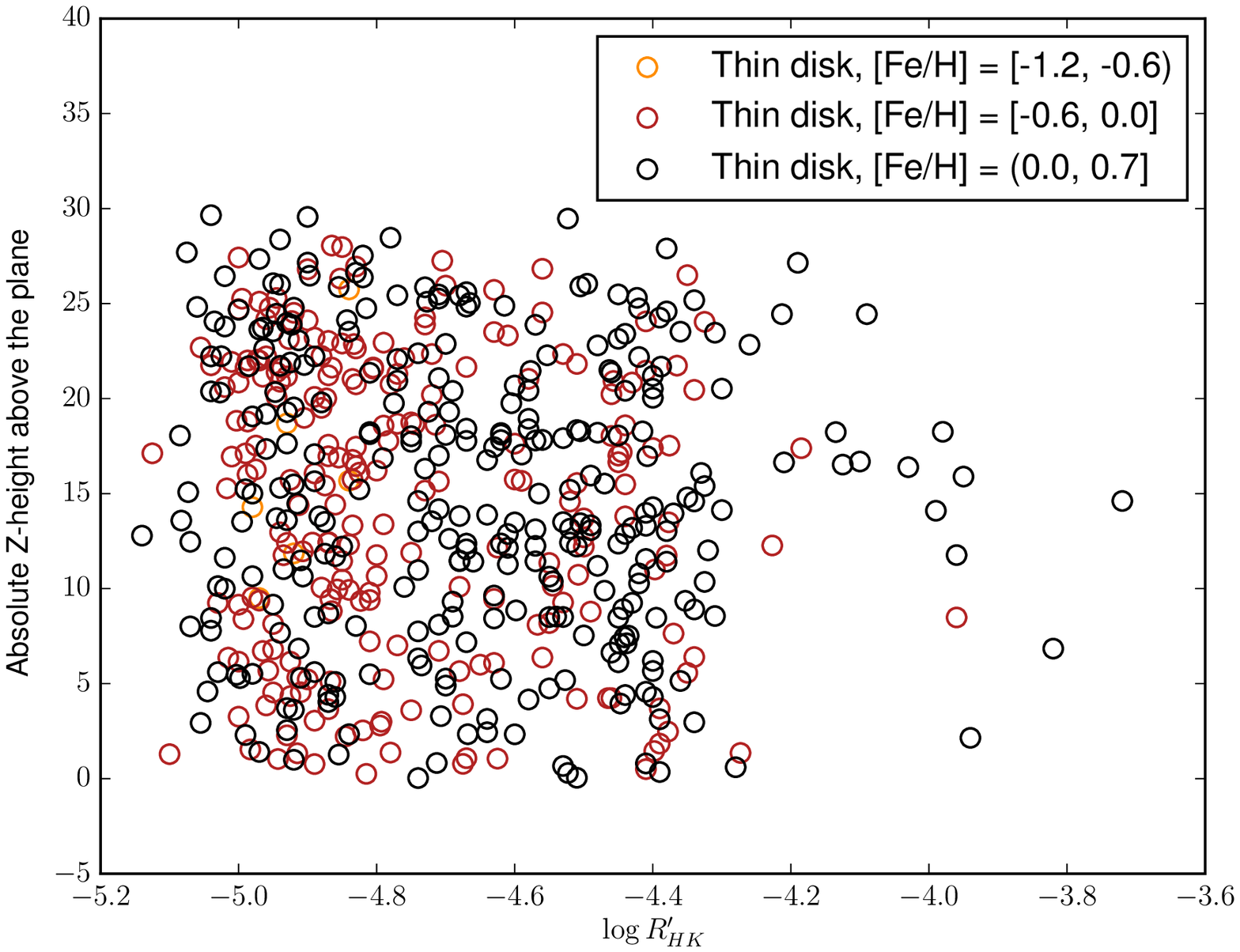}} 
\subfloat{\includegraphics[width = 3in]{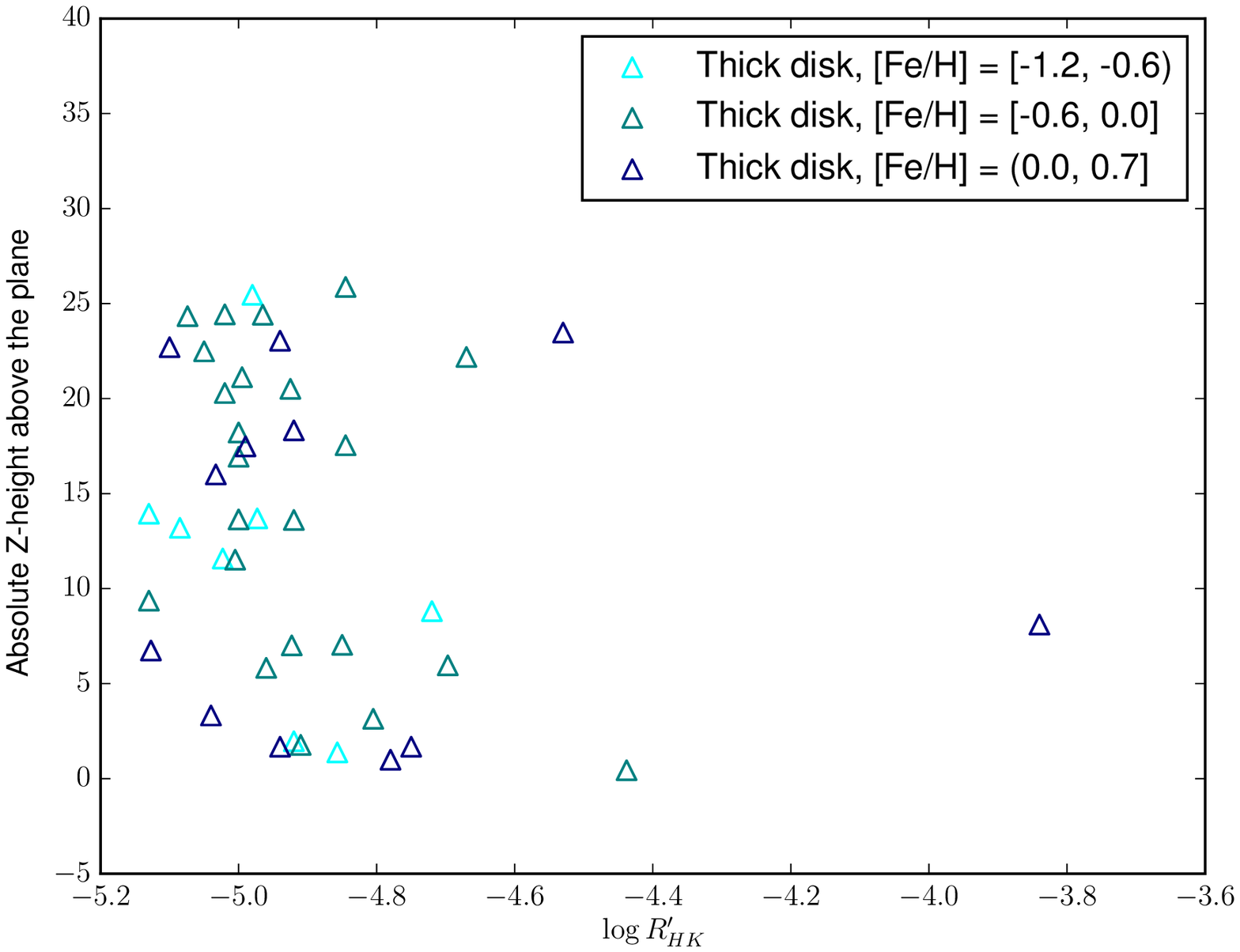}}\\
\caption{Stellar activity with respect to the absolute Z-height above the plane for [Fe/H], where stars in the thin disk are shown on the left while stars in the thick disk are on the right.} 
\label{RHKvsZ}
\end{figure*}

\begin{figure*}
\center
\subfloat{\includegraphics[width = 3.5in]{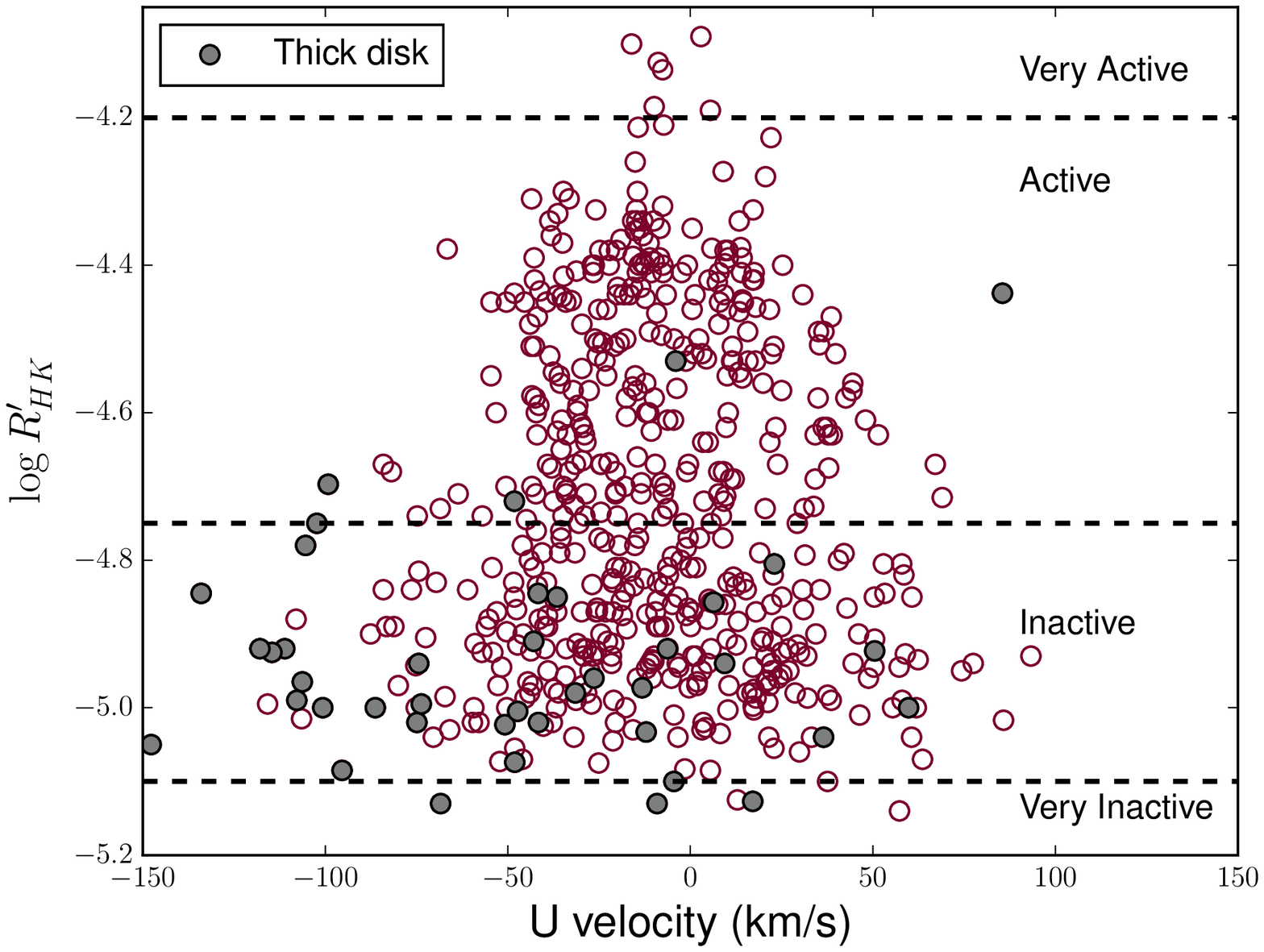}} 
\subfloat{\includegraphics[width = 3.5in]{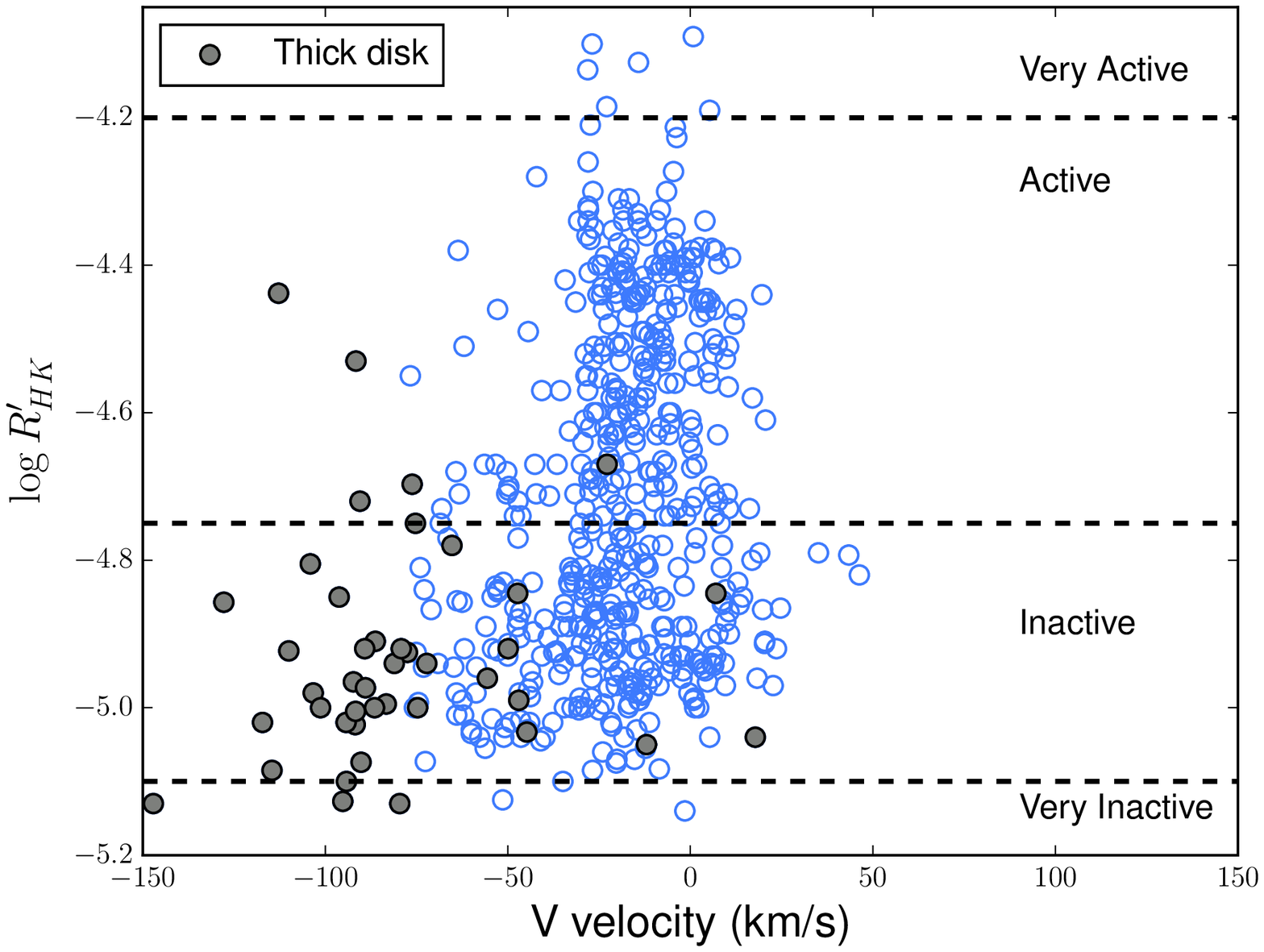}}\\
\subfloat{\includegraphics[width = 3.5in]{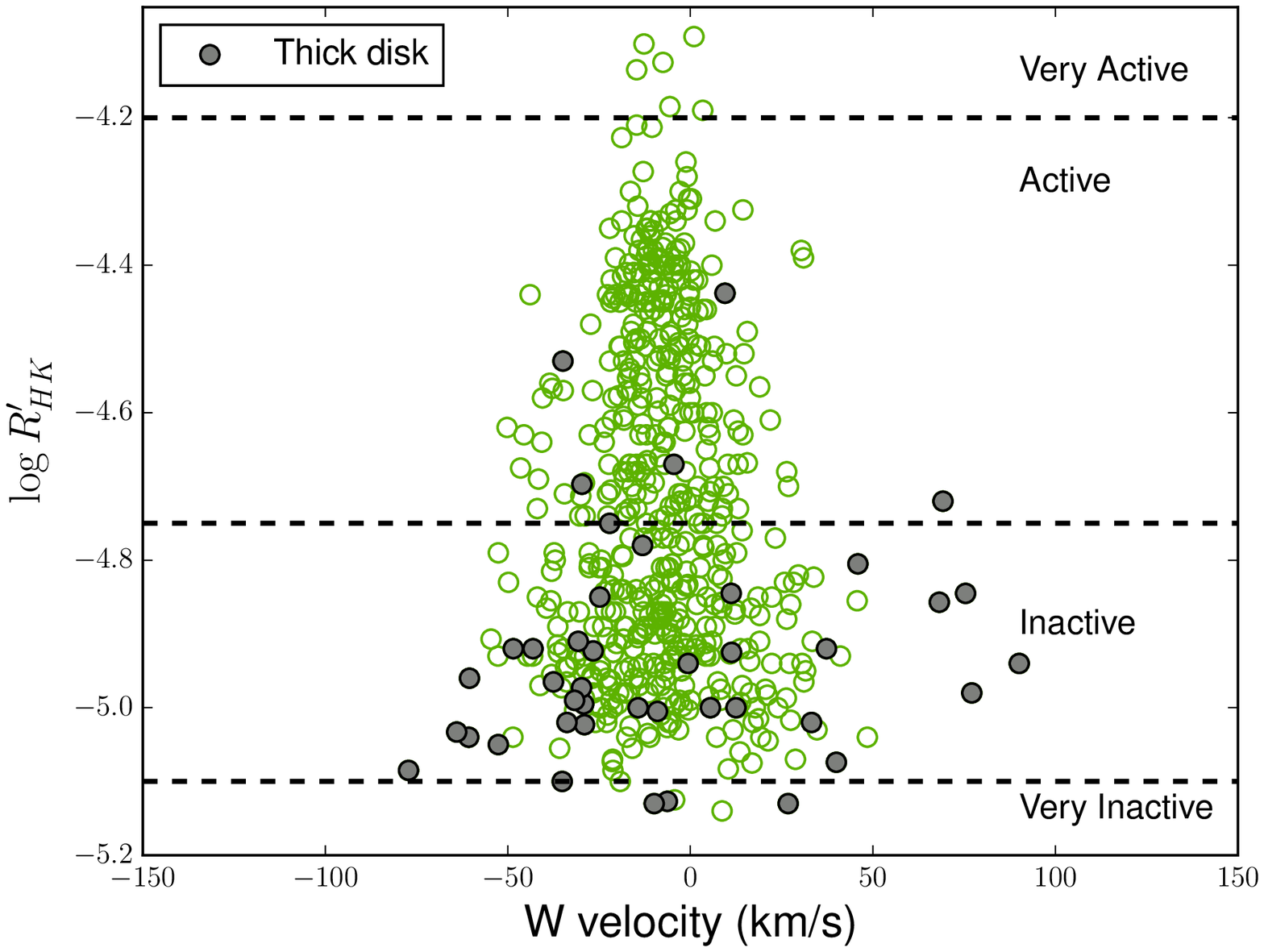}}
\subfloat{\includegraphics[width = 3.5in]{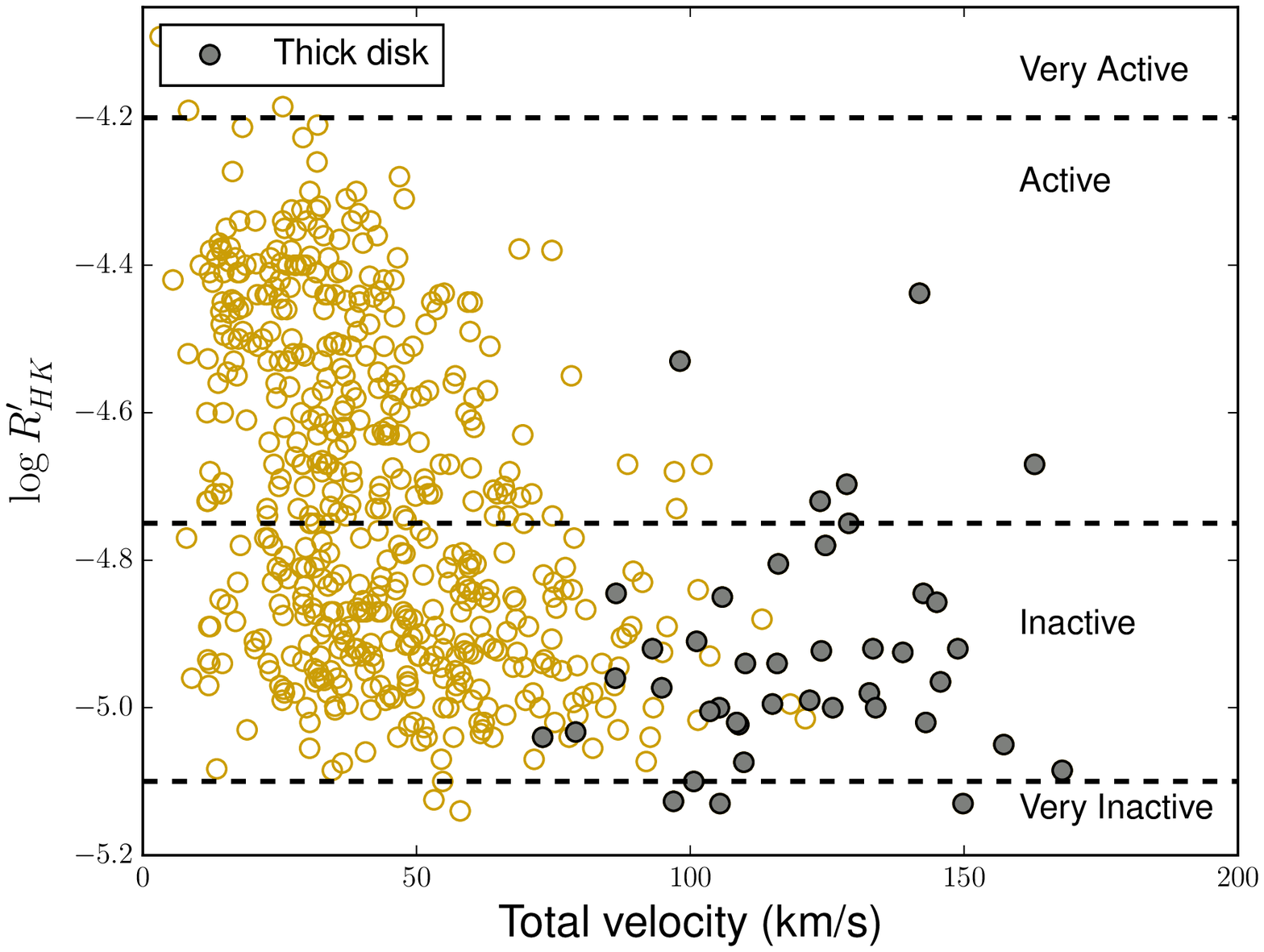}} \\
\caption{ The chromospheric activity
indicator $\log R_{HK}^{\prime}$, based on the Ca II H and K emission lines,
is plotted versus the U, V, and W velocity components of space motion as well as the total
space velocity for stars in the CATSUP sample. Thick disk stars are depicted
with filled symbols, thin disk stars with open circles. The four stellar
activity groupings defined by \citet{Henry96} are indicated by the
horizontal dashed lines and labelled accordingly.} 
\label{UVW-logRHK}
\end{figure*}

\begin{table*}
\caption{A summary of the velocity components of CATSUP stars
with different activity levels}
\begin{center}
\begin{tabular}{c|cccc}
\hline
\hline
 & Very Active   &    Active   &    Inactive  &   Very Inactive \\
 $\log R_{HK}^{\prime}$ & $x > -4.2$ & $-4.75 <  x \le -4.2$  & $-5.1 < x \le -4.75$ &  $x \le -5.1$ \\
\hline 
mean U   &   -12.40        &        -10.83      &    -13.69      &     1.96\\
std U     &     22.02         &         29.22      &    42.16       &    41.19\\
mean V    &   -22.81      &          -16.42      &     -28.81     & -74.96\\
std V      &     17.25        &          18.91        &    29.99     &    48.14\\
mean W  &   -13.86       &         -8.76            & -7.04       &    3.00\\
std W     &     13.85         &       13.86            & 23.64      &   13.45\\
\hline
\hline
\end{tabular}\label{tab.activity}
\end{center}
\end{table*}

The CATSUP dataset combines a variety of properties for stars within 30 pc, such that individual stars as well as the solar neighborhood can be better characterized. To take advantage of the assorted available properties, we have plotted the \ion{Ca}{2} H and K emission index $\log R^{\prime}_{\rm HK}$ with respect to both B-V color in Fig. \ref{activity}. We have color-coded the stars to show the likely disk component of origin, where orange is thin disk and blue is thick disk, based on their kinematics per \citep{Bensby03, Hinkel14}. The thick disk stars are mainly concentrated around $\log R^{\prime}_{\rm HK}$ of $\sim$ -5.0 $\pm$ 0.2. In other words, they have emission indices comparable to the Sun --  they are mostly very low activity. Along these lines, we considered whether the height above/below the Galactic plane (Z) and [Fe/H] correlate with a stellar activity index (that is sensitive to age). We did not see any trends with respect to
$\log R^{\prime}_{\rm HK}$ vs. $|$Z$|$, or with respect to a multitude of abundances within both thin- and thick-disk stars, see Fig. \ref{RHKvsZ}. The conclusion that we can draw is that CATSUP stars encompass the full range of Z-height values in both the thin and thick disk subsets regardless of their level of Ca II H and K stellar activity indices (at least for $\log R^{\prime}_{\rm HK}$ $<$ -4.8 or so).

The thick disk stars typically have $\log R_{HK}^{\prime} < -4.8$, and so fall
in the low-activity grouping below the Vaughan-Preston gap \citep{Vaughan80} in plots of Ca II
H and K activity versus B-V. As such the thick disk stars populate the peak at
$\log R_{HK}^{\prime} \sim -4.9$ in Fig. \ref{freqRHK}. If the thick disk stars are removed
from the histogram in Fig. \ref{freqRHK}, the resulting distribution for thin disk stars
still shows a pronounced local maximum near $-4.9$. Thus a high incidence of
inactive stars is a property of the thin disk as well as the thick disk.

Looking at the two populations of stars in Fig. \ref{activity}, there are 536 stars from the thin disk while 43 are from the thick disk ($\sim$8\%). To better understand these two stellar populations, we utilize a two-sample Kolmogorov-Smirnov (KS) test which analyzes whether the two samples are drawn from the same distribution. If the p-value is below a certain significance level, typically 0.05, then the null hypothesis (that the two samples are from the same distribution) is rejected. Note, the p-value is not the probability of the null hypothesis being true or false. In other words, a p-value $>$ 0.05 does not mean that the two samples are similar, it merely states that there was no evidence to show that the two samples were significantly different. This point is subtle and often misinterpreted within the literature. A two-sample KS test of the $\log R^{\prime}_{\rm HK}$ values for thin- and thick-disk stars yields a p-value = 2.02 $\times$ 10$^{-7}$ -- meaning that the two samples are statistically different. 

We see in Fig. \ref{activity} that thick-disk stars have predominantly low values of $\log R^{\prime}_{\rm HK}$, typically $\log R^{\prime}_{\rm HK} <$ -4.8. For all stars with $\log R^{\prime}_{\rm HK}$ $<$ -4.8, there are 233 stars from the thin disk and 35 from the thick disk ($\sim$15\%). Doing a two-sample KS test on the $\log R^{\prime}_{\rm HK}$ values for thin- and thick-disk stars where $\log R^{\prime}_{\rm HK}$ $<$ -4.8 gives p = 0.006. Finally, we did a two-sample KS test for the $\log R^{\prime}_{\rm HK}$ values for those stars in the thick disk and with $\log R^{\prime}_{\rm HK}$ $<$ -4.8 (total 35 stars) with respect to the entire sample of thin-disk stars (total 536 stars) -- the p-value is 2.82 $\times$ 10$^{-11}$. 

\citet{Vaughan80} searched for kinematic differences among some 185
dwarfs with different levels of Ca II H and K activity. They found that for
dwarfs of a given spectral type, groupings according to strong H and K emission
(high activity) have smaller dispersions in the component of space motion
perpendicular to the Galactic plane. Thus the chromospherically younger stars
in their sample have different mean space motions than chromospherically older
stars. \citet{Soderblom90} extended this work with a larger sample of
chromospherically active solar-like, K, and M dwarfs, finding them to have
kinematics consistent with a young population of age around 0.5-2 Gyr. The age
dependence of space motion provides a tool for studying the dynamical evolution
of the Galactic disk \citep[e.g.][]{Wielen74}. As such any correlations between
stellar kinematics and stellar activity are worth searching for among the
CATSUP stars. \citet{Jenkins11} have published a very thorough study
of the correlations for solar-type dwarfs and subgiants, highlighting the
utility of this approach. 

The distributions of the (U,V,W) components of space motion of the CATSUP stars
are shown in Fig. \ref{UVW-logRHK}, with some summary characteristics being listed in Table
\ref{tab.activity}. In the table the CATSUP stars have been divided into the four stellar
activity groupings defined by \citet{Henry96}, as also adopted in Section
\ref{uv}, and for each grouping the mean value of each velocity component is listed
along with the standard deviation. We leave the ``very inactive'' stars out
of the discussion because of their small number within the CATSUP sample.
In terms of the W component of velocity perpendicular to the Galactic plane,
the velocity dispersion among the inactive stars is notably greater than among
the active and very-active dwarfs. This is also the case for the standard
deviations in the U and V components of motion. Thus overall the inactive stars
evince a greater dispersion in all three velocity components than the active
stars. This trend is partly but not entirely driven by the thick disk dwarfs
within CATSUP. Inspection of Fig. \ref{UVW-logRHK} shows that these trends exist
even within the thin disk population alone. The upper panels of Fig.~\ref{UVW-logRHK}
illustrate the offsets in mean U and V velocity between the thick and thin
disk stars. Thus the CATSUP sample verifies and extends the early results of
\citet{Vaughan80}, and is consistent with the findings of \citet{Soderblom90}
and \citet[][see their Fig. 16 to which our Fig. \ref{UVW-logRHK} is
analogous]{Jenkins11}.

Ultimately, it seems that the thick-disk stars have predominantly low activity, and higher velocity dispersions on average toward or away-from the Galactic plane. The CATSUP set of thick-disk stars has a distinctly different distribution of chromospheric activity than the CATSUP thin-disk stars. We note as a caveat, however, that the number of thick disk stars in our sample is roughly an order of magnitude smaller than the thin-disk stars.

\section{Summary}

We have assembled a dataset of stellar properties for 951 FGK-type stars within 30 pc of the Sun. Beginning with the Gaia TGAS subset of astrometric data, we have combined information regarding multiplicity within stellar systems (ExoCat), stellar abundance measurements (Hypatia), standardized spectral types, \ion{Ca}{2} H and K stellar activity indices, NUV and FUV photometry from GALEX, and X-ray fluxes and luminosities from ROSAT, XMM, and Chandra. The aim of this project was to collate a wide variety of data for nearby stars such that they could be more easily characterized. The information available in CATSUP can be utilized for the direct sample or act as a proxy for similar stars, in order to better understand the overall trends within solar neighborhood stars as well as stars that host exoplanets. CATSUP was compiled 
in anticipation of upcoming exoplanet surveys such as TESS, CHEOPS, and WFIRST. 

While we included data currently available within the literature, we also presented new stellar information. We explored the GALEX UV data and found that the FUV and NUV flux correlated strongly with effective temperature as the photospheres of hotter stars have bluer peaks in their Planck functions. At temperatures below 5500 K, the data show a much large range at a given \teff due to higher contributions to the FUV emission from the more variable stellar corona and high chromosphere compared with hotter stars (see Fig.  \ref{fuvnuv}, left). At higher temperatures, the FUV flux is more linear with \teff since it is dominated by the temperature dependent stellar photosphere. In general, the NUV data show a large range at a given effective temperature. When combined into a flux density ratio in the FUV and NUV band passes per Fig. \ref{fuvnuv} (right) the influence of the stellar corona/transition region/chromosphere over the photosphere at low temperatures is again present. Additionally, we analyzed how $\log R^{\prime}_{\rm HK}$ activity effected stellar spectral energy distributions, such that more active stars had bluer FUV-V colors.

\begin{figure}
\begin{center}
 \centerline{\includegraphics[width=10cm]{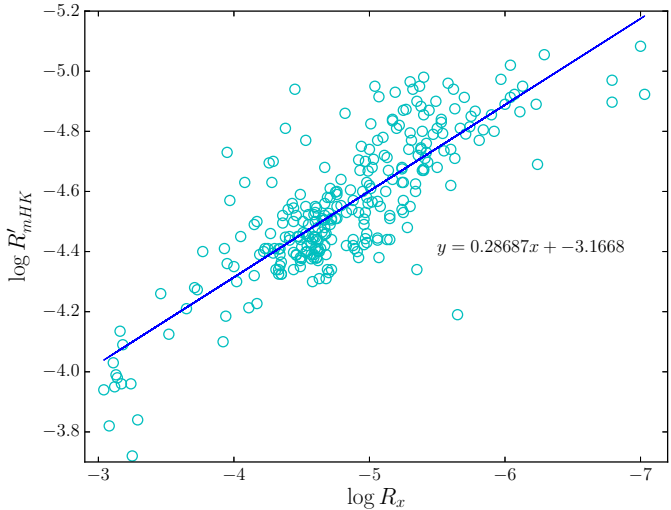}}
\end{center}
  \caption{The Ca II H and K indices with respect to the fractional X-ray luminosity, where $\log R_x$ = $\log \,( L_x/L_{bol} )$. A linear regression line is overlaid in blue, with the equation given. The scatter is usual for a sample of stars such as CATSUP.}
  \label{actvsact}
    \vspace{3mm}
\end{figure}

X-ray data from multiple missions were combined per a new methodology that allowed all available X-ray information to be utlized for stellar characterization. The X-ray data were investigated with respect to effective temperature in Fig. \ref{xrayteff}. The majority of the stars were found to fall between the solar coronal activity and the X-ray saturation limit, where the latter was exemplified by a relatively constant $\log R_x$ across spectral types. Overall, the smaller, cool stars were found to have the highest fractional X-ray luminosity or coronal activity indicator. We compared the coronal activity ($\log R_x$) and chromospheric activity ($\log R^{\prime}_{\rm HK}$) in the same manner as \citet{Mamajek08}, see Fig. \ref{actvsact}. We found a similar correlation with respect to a linear trend, where the overlaid line is defined as $y = 0.28687 x - 3.1668$, and range in data. 

Finally, we examined the correlation between stars in the CATSUP sample that are likely to have originated from the thick disk (see Section \ref{hypatia}) and the Ca II H and K index $\log R^{\prime}_{\rm HK}$. We found that the thick-disk stars had a chromospheric activity that was preferentially $\log R^{\prime}_{\rm HK}$ $<$ -4.8 in Fig \ref{activity}. Compared with the thin-disk stars, these kinematic stellar sub-groups within CATSUP were statistically different per a two-sample KS test. When analyzing $\log R^{\prime}_{\rm HK}$ with respect to UVW galactic velocity, the lower activity stars had greater dispersion in the three individual galactic velocity components, a trend which is not wholly attributable to the presence of thick-disk stars. 

While there are a number of additional trends to be found between the 951 CATSUP stars, we leave this task to future papers either by the ASU NExSS team or other colleagues. The properties within CATSUP were strategically combined in order to maximize characterization of nearby main-sequence stars. It is our goal that by knowing more about stars from both a physical and chemical perspective, we will enable a greater understanding of the solar neighborhood. Additionally, we hope that CATSUP will help inform either target selection or follow-up observations for the TESS, CHEOPS, and WFIRST missions.

\section*{Acknowledgements}
The results reported herein benefited from collaborations and/or
information exchange within NASA's Nexus for Exoplanet System Science
(NExSS) research coordination network sponsored by NASA's Science
Mission Directorate.
NRH would like to thank CHW3 and Tatertot, who goes well with CATSUP. 
NRH and GES acknowledge the support of the
Vanderbilt Office of the Provost through the Vanderbilt Initiative in
Data-intensive Astrophysics (VIDA) fellowship.
EO and ES appreciate support from NASA/Habitable Worlds grant NNX16AB62G.
GHS wishes to acknowledge the support of NSF Award 1517791.
Part of this research was carried out at the Jet Propulsion
Laboratory, California Institute of Technology, under a contract with
NASA. This document does not contain any export controlled technical data
(URS268608).
Finally, this research has made use of the Vizier catalogue access tool, 
CDS, Strasbourg, France as well as Filtergraph (for the HR diagram), an online data visualization tool 
developed at Vanderbilt University through the Vanderbilt Initiative in Data-intensive Astrophysics (VIDA).

\newpage
\clearpage

\tiny
\begin{turnpage}
\begin{deluxetable}
{|p{2.0cm}p{4.0cm}p{1.0cm}p{1.0cm}p{1.7cm}p{1.3cm}p{1.3cm}p{2.0cm}p{1.8cm}p{1.0cm}p{1.0cm}|}
\tablecaption{\label{update1} Hypatia 2.1 Update}
\tablehead{
\colhead{Catalog} &
\colhead{Telescope} &
\colhead{Resolution} &
\colhead{S/N} &
\colhead{$\lambda$ Range} &
\colhead{Stellar} &
\colhead{Eq.} &
\colhead{CoG or} &
\colhead{Solar} &
\colhead{Num. of } &
\colhead{Stars in } \\
\colhead{} &
\colhead{} &
\colhead{($\Delta \lambda / \lambda$)} &
\colhead{} &
\colhead{(\aaa)} &
\colhead{Atmo} &
\colhead{Width} &
\colhead{SF} &
\colhead{Scale} &
\colhead{FeI/II lines} &
\colhead{Hypatia}\\
}
\startdata
\hline
\hline
\citet{Adibekyan16} & HARPS (3.6 m ESO telescope,  La Silla,  Chile) and UVES (8m VLT / UT2 telescope,  La Silla,  Chile) &  & 200-2300 &  & ATLAS9 per \citet{Kurucz1993} & ARES & MOOG per \citet{Sneden73} & \citet{Anders:1989p3165} & 250 / 40 & 39 \\ 
\citet{Battistini16} & FEROS on the ESO 1.5-m and 2.2-m telescopes and MIKE on the Magellan Clay telescope; UVES on the ESO Very Large Telescope & 42000 - 65000 & $>$ 200 & 3500-9500 & MARCS per \citep{Gustafsson:1975p4658} & IRAF splot & Uppsala EQWIDTH & differential analysis & 226 / 36 & 471\\ 
\citet{Baumann10} & Robert G. Tull coude spectrograph on the 2.7 m Harlan Smith telescope; MIKE spectrograph on the 6.5 m Magellan Clay telescope; HARPS spectrograph on the 3.6 m ESO telescope  & 45000 - 110000 & $>$ 200 & 4500-7800  /  3350-9500 / 4445-8294 & ATLAS9 per \citet{Kurucz1993} & IRAF splot & MOOG per \citet{Sneden73} & differential analysis & 34 / 11  & 117\\ 
\citet{Brewer16} & HIRES spectrograph at the Keck I Telescope & 70000 & $>$200 & 5164-7800 & ATLAS per \citet{Kurucz1993} & spectral fitting & spectral fitting & differential analysis & 600 / 300 & 894\\ 
\citet{Chen01} & 2.16m telescope at Beijing Astronomical Observatory (BAO) with the Coude Echelle Spectrograph (CES) & 40000 & $>$150 & 5500-9000 & MARCS per \citep{Gustafsson:1975p4658} & SPECTRUM & Uppsala EQWIDTH & \citet{Anders:1989p3165} & 142 / 8 &  143\\ 
\citet{daSilva15} & Observatoire de Haute-Provence (OHP) using the ELODIE spectrograph & 42000 & $>$ 200 & 3895-6815  & ATLAS9 per \citet{Kurucz1993} & ARES & MOOG per \citet{Sneden73} & differential analysis & 72 / 12  &  304\\
\hline
\tablenotetext{*}{Telescope/spectrograph information and the
techniques for determining abundances as given by the recently added
literature sources (with more than 20 stars added) into {\it
Hypatia} 2.1. Please see \citet{Hinkel14, Hinkel16} for more
details. In the header, ``S/N" is signal-to-noise report by the literature source, ``$\lambda$ Range" is the wavelength coverage, ``Stellar Atmo" is the stellar atmospheric model, ``Eq. Width" is the package used to determine the equivalent width, ``CoG or SF" designates whether the group used a curve-of-growth or spectral fitting technique where the package is specified in the former case, the ``Solar Scale" is the solar normalization used by that group (differential analysis is cited where applicable), and ``Num. of Fe I/II lines" lists the number of Fe I and Fe II lines. }
\enddata
\end{deluxetable}
\clearpage
\end{turnpage}

\tiny
\LongTables
\begin{landscape}
\begin{deluxetable} {|p{2.0cm}p{4.0cm}p{1.0cm}p{1.0cm}p{1.7cm}p{1.3cm}p{1.3cm}p{2.0cm}p{1.8cm}p{0.8cm}p{1.0cm}|}
\tablecaption{\label{update2} Hypatia 2.1 Update}
\tablehead{
\colhead{Catalog} &
\colhead{Telescope} &
\colhead{Resolution} &
\colhead{S/N} &
\colhead{$\lambda$ Range} &
\colhead{Stellar} &
\colhead{Eq.} &
\colhead{CoG or} &
\colhead{Solar} &
\colhead{Num. of } &
\colhead{Stars in } \\
\colhead{} &
\colhead{} &
\colhead{($\Delta \lambda / \lambda$)} &
\colhead{} &
\colhead{(\aaa)} &
\colhead{Atmo} &
\colhead{Width} &
\colhead{SF} &
\colhead{Scale} &
\colhead{FeI/II lines} &
\colhead{Hypatia}\\
}
\startdata
\hline
\hline
\citet{DelgadoMena14, DelgadoMena15} & HARPS at 3.6-m ESO La Silla Observatory (Chile); UVES at 8.2-m Kueyen UT2 (VLT); FEROS at 2.2-m ESO/MPI telescope; SARG 3.5-m TNG; FIES at 2.6-m Nordic Optical Telescope; SOPHIE at 1.93-m OHP; CORALIE at 1.2-m Euler Swiss telescope; and UES at 4.2-m William Herschel Telescope  & 100000 / 115000 / 48000 / 57000-86000 / 67000 / 75000 / 50000 / 55000 & 55\% of the spectra $>$ 200 & 3800-7000 / 3000-4800; 4800-6800 / 3600-9200 / 5100-10100 / 3700-7300 / 3820-6930 / 3800-6800 / 4600-7800 & ATLAS9 per \citet{Kurucz1993} & IRAF splot & MOOG2010 per \citet{Sneden73} & \citet{Anders:1989p3165} & N/A & 137 / 287\\ 
\citet{Gonzalez01, Gonzalez10a, Gonzalez10b} & 2.7m telescope at McDonald using the 2dcoude spectrometer and 4m Blanco Telescope at Cerro Tololo Inter-American Observatory (CTIO) & $>$ 59000 / 35000 & 195-620 & 3700-10000   /  5850-8950  & ATLAS9 per \citet{Kurucz1993} & Uppsala EQWIDTH & MOOG per \citet{Sneden73} & their own & 64 / 11 & 20 / 124\\ 
\citet{Gonzalez14, Gonzalez15} & McDonald Observatory 2.1-m Otto Struve telescope and Sandiford spectrograph & 53000 & 300-350 &  & ATLAS per \citet{Kurucz1993} & DAOSPEC & MOOG per \citet{Sneden73} & differential analysis & 45-55 / 6-8 & 37 / 30\\ 
\citet{Ghezzi10b} & FEROS on the MPG/ESO-2.20 m telescope & 48000 & $>$ 200 & 3560-9200  & ODFNEW / ATLAS9 per \citet{Kurucz1993} & ARES & MOOG per \citet{Sneden73} &  & 27 / 12 &  149\\ 
\citet{Israelian04, Israelian09} & 4.2 m WHT/UES (La Palma); the 3.5 m TNG/SARG (La Palma); the 1.52 m ESO (La Silla) and the 1.2 m Swiss/CORALIE (La Silla) & 55000 / 57000 / 50000 / 50000 & 150-350 & 3800-6800  & ATLAS9 per \citet{Kurucz1993} per \citet{Kurucz1993} & IRAF splot & MOOG per \citet{Sneden73} & \citet{Anders:1989p3165} & 40 / 7 & 61 / 127 \\ 
\citet{Lambert91} & W. J. McDonald Observatory with the coude spectrographs of the 2.1-m and 2.7-m reflectors &  & $>$ 150 &  & ATLAS per \citet{Kurucz1993} & Uppsala EQWIDTH & Uppsala EQWIDTH & \citet{Anders:1989p3165} & &  61\\ 
\citet{Lambert04} & 2.7 m telescope at McDonald using the 2dcoude spectrometer & 60000 & 100-200 & 3500-9000  & ATLAS9 per \citet{Kurucz1993} & IRAF splot & MOOG per \citet{Sneden73} & \citet{Reddy:2003p1354} & 54 / 9 &  103\\ 
\citet{Liu14} & High Dispersion Echelle Spectrograph at Okayama Astrophysical Observatory (OAO); which was equipped at the coude focus of the 1.88 m telescope & 67000 & $>$ 200 & 5000-6200; 4000-7540 & MAFAGS & spectral synthesis with IDL/Fortran SIU software \citep{Reetz91} & spectral fitting & differential analysis & & 83\\ 
\citet{LopezValdivia17} &	2.1-m telescope of the Observatorio Astrofsico Guillermo Haro, located in Mexico, using the Cananea High- resolution Spectrograph (CanHiS)&	80000&	100	&centered at 5005, 5890, 6310 and 6710\aaa	& ATLAS12 per \citet{Kurucz1993} &	their own analysis&	MOOG per \citet{Sneden73}	&their own off Vesta&	18 / 0 & 22\\
\hline
\tablenotetext{*}{Telescope/spectrograph information and the
techniques for determining abundances as given by the recently added
literature sources (with more than 20 stars added) into {\it
Hypatia} 2.1. Please see \citet{Hinkel14, Hinkel16} for more
details. In the header, ``S/N" is signal-to-noise report by the literature source, ``$\lambda$ Range" is the wavelength coverage, ``Stellar Atmo" is the stellar atmospheric model, ``Eq. Width" is the package used to determine the equivalent width, ``CoG or SF" designates whether the group used a curve-of-growth or spectral fitting technique where the package is specified in the former case, the ``Solar Scale" is the solar normalization used by that group (differential analysis is cited where applicable), and ``Num. of Fe I/II lines" lists the number of Fe I and Fe II lines. }

\enddata
\end{deluxetable}
\clearpage
\end{landscape}

\tiny
\LongTables
\begin{landscape}
\begin{deluxetable} {|p{2.0cm}p{4.0cm}p{1.0cm}p{1.0cm}p{1.7cm}p{1.3cm}p{1.3cm}p{2.0cm}p{1.8cm}p{1.0cm}p{1.0cm}|}
\tablecaption{\label{update3} Hypatia 2.1 Update}
\tablehead{
\colhead{Catalog} &
\colhead{Telescope} &
\colhead{Resolution} &
\colhead{S/N} &
\colhead{$\lambda$ Range} &
\colhead{Stellar} &
\colhead{Eq.} &
\colhead{CoG or} &
\colhead{Solar} &
\colhead{Num. of } &
\colhead{Stars in } \\
\colhead{} &
\colhead{} &
\colhead{($\Delta \lambda / \lambda$)} &
\colhead{} &
\colhead{(\aaa)} &
\colhead{Atmo} &
\colhead{Width} &
\colhead{SF} &
\colhead{Scale} &
\colhead{FeI/II lines} &
\colhead{Hypatia}\\
}
\startdata
\hline
\hline
\citet{Luck06} & 2.1m telescope at McDonald using the CASPEC & 60000 & $>$ 150 & 4840-7000  & MARCS per \citep{Gustafsson:1975p4658}75 & spectral fitting & spectral fitting & differential analysis & 450 / 25 &  194\\ 
\citet{Luck07} & 2.1m telescope at McDonald using the CASPEC & 60000 & $>$ 150 & 4840-7000  & MARCS per \citep{Gustafsson:1975p4658} & LINES & MOOG per \citet{Sneden73} & differential analysis & 332 / 18 &  296\\ 
\citet{Luck15, Luck17} & McDonald Observatory using the 2.1m Struve Telescope and the Sandiford Cassegrain Echelle Spectrograph & 60000 & $>$ 150 & 4840-7000  & MARCS per \citep{Gustafsson:1975p4658} & LINES & MOOG per \citet{Sneden73} &  differential analysis & 500 / 30-50 &  904 / 967\\ 
\citet{Mahdi16} & ELODIE was on the 1.93 m telescope at Observatoire de Haute- Provence (OHP) & 42000 & $>$ 70 & 4000-6800  & MARCS per \citep{Gustafsson:1975p4658} & iSpec per \citet{Blanco14} & iSpec per \citet{Blanco14} & differential analysis & 189 \\ 
\citet{Maldonado15, Maldonado16} & HERMES spectrograph at the MERCATOR (1.2 m) telescope at La Palma observatory and FIES at the Nordic Optical Telescope (2.56 m) & 85000 / 67000 & 90-340 / 75-480 & 3800-9000  / 3640-7360  & ATLAS9 per \citet{Kurucz1993} & WIDTH9 per \citep{Kurucz1993} & WIDTH9 per \citep{Kurucz1993} & differential analysis) & 263 / 36 & 251 / 142\\ 
\citet{Mallik03} & coude echelle spectrograph at the 102 cm telescope at the Vainu Bappu Observatory at Kavalur &  &  &  & MARCS per \citep{Gustafsson:1975p4658} &  & MOOG per \citet{Sneden73} & differential analysis & &  71\\ 
\citet{Mishenina12} & 1.93m telescope at OHP using ELODIE & 42000 & 100-350 & 3850-6800  & ATLAS9 per \citet{Kurucz1993} & WIDTH9 per \citep{Kurucz1993} & N/A & differential analysis & N/A &  59\\ 
\citet{Mishenina16} & 1.93m telescope at OHP using ELODIE & 42000 & 100-350 & 4400-6800  & ATLAS9 per \citet{Kurucz1993} & WIDTH9 per \citep{Kurucz1993} & N/A & differential analysis & N/A & 196\\ 
\citet{Nissen13} & HARPS at 3.6-m ESO La Silla Observatory (Chile); FEROS at 2.2-m ESO/MPI telescope & 115000 / 48000 & 250-1000 / 200-300 & 3800-6900 / 3500-9200 & MARCS per \citep{Gustafsson:1975p4658} & IRAF splot & Uppsala EQWIDTH & differential analysis & N/A &  33\\ 
\citet{Nissen16} & HARPS (3.6 m ESO telescope,  La Silla,  Chile) & 115000 & $> $600 & 3800-6900  & MARCS per \citep{Gustafsson:1975p4658} & IRAF splot & Uppsala EQWIDTH & differential analysis & 47 / 9 \\ 
\citet{Notsu17} & High Dispersion Echelle Spectrographattached at the 1.88-m reflector of Okayama Astrophysical Observatory (OAO) & 59000 &  & 5600-9100 and 4300-7700 & ATLAS9 per \citet{Kurucz1993} & WIDTH9 per \citep{Kurucz1993} & SPSHOW (in the SPTOOL software developed by Y. Takeda; unpublished) &  \citet{Anders:1989p3165} & 160 / 20 &  36\\ 
\citet{Pagano17} & Magellan Inamori Kyocera Echelle (MIKE) spectrograph on the 6.5 meter Magellan II telescope & 50000 & 150-300 & 4700-7100 & ATLAS9 per \citet{Kurucz1993}  & ARES/IRAF {\sc SPLOT} & MOOG14 per \citet{Sneden73} & \citet{Asplund:2009p3251} & 69 / 14 &  508\\ 
\citet{Ramirez12a} & Tull coude spectrograph on the 2.7m Harlan J. Smith Telescope at McDonald Observatory; MIKE spectrograph on the 6.5m telescope at Las Campanas Observatory & 60000 (both) & $>$200 & N/A & MARCS per \citep{Gustafsson:1975p4658} & IRAF splot & MOOG per \citet{Sneden73} & differential analysis & N/A &  514\\ 
\hline
\tablenotetext{*}{Telescope/spectrograph information and the
techniques for determining abundances as given by the recently added
literature sources (with more than 20 stars added) into {\it
Hypatia} 2.1. Please see \citet{Hinkel14, Hinkel16} for more
details. In the header, ``S/N" is signal-to-noise report by the literature source, ``$\lambda$ Range" is the wavelength coverage, ``Stellar Atmo" is the stellar atmospheric model, ``Eq. Width" is the package used to determine the equivalent width, ``CoG or SF" designates whether the group used a curve-of-growth or spectral fitting technique where the package is specified in the former case, the ``Solar Scale" is the solar normalization used by that group (differential analysis is cited where applicable), and ``Num. of Fe I/II lines" lists the number of Fe I and Fe II lines. }

\enddata
\end{deluxetable}
\clearpage
\end{landscape}

\tiny
\LongTables
\begin{landscape}
\begin{deluxetable} {|p{2.0cm}p{4.0cm}p{1.0cm}p{1.0cm}p{1.7cm}p{1.3cm}p{1.3cm}p{2.0cm}p{1.8cm}p{1.0cm}p{1.0cm}|}
\tablecaption{\label{update4} Hypatia 2.1 Update}
\tablehead{
\colhead{Catalog} &
\colhead{Telescope} &
\colhead{Resolution} &
\colhead{S/N} &
\colhead{$\lambda$ Range} &
\colhead{Stellar} &
\colhead{Eq.} &
\colhead{CoG or} &
\colhead{Solar} &
\colhead{Num. of } &
\colhead{Stars in } \\
\colhead{} &
\colhead{} &
\colhead{($\Delta \lambda / \lambda$)} &
\colhead{} &
\colhead{(\aaa)} &
\colhead{Atmo} &
\colhead{Width} &
\colhead{SF} &
\colhead{Scale} &
\colhead{FeI/II lines} &
\colhead{Hypatia}\\
}
\startdata
\hline
\hline
\citet{Ramirez13} & TS2/McD (R. G. Tull Coude spectrograph; 2.7 m Telescope at McDonald Observatory); HRS/HET (High Resolution Spectrograph; 9.2 m Hobby-Eberly Telescope); UVES/VLT (UV-Visual Echelle Spectrograph; 8 m Very Large Telescope); and FEROS/ESO (Fiber-feb Extended Range Optical Spectrograph; ESO 1.52-m Telescope) & 60000 / 120000 / 80000 / 45000 & $>$ 100 &  & MARCS per \citep{Gustafsson:1975p4658} & IRAF splot & MOOG per \citet{Sneden73} & differential analysis  & &  794\\ 
\citet{SuarezAndres16} & 3.6m telescope at ESO equipped with High Accuracy Radial Velocity Planet Searcher (HARPS) using CORALIE & 110000 & 70-2000 & 3800-6900  & ATLAS9 per \citet{Kurucz1993} & ARES & MOOG per \citet{Sneden73} & \citet{Anders:1989p3165} & 263 / 36 &  1077\\ 
\citet{Takeda05Li, Takeda10} & 1.88m telescope at Okayama Astrophysical Observatory (OAO) using the High Dispersion Echelle Spectrograph (HIDES) & 70000 & 200 & 5800-7000 & ATLAS9 per \citet{Kurucz1993} & WIDTH9 per \citep{Kurucz1993} & SPSHOW (in the SPTOOL software developed by Y. Takeda; unpublished) & \citet{Anders:1989p3165} & 160 / 20 &  93 / 83\\ 
\citet{Trevisan14} & 1.52m telescope at ESO using FEROS & 48000 & 100 & 3560-9200  & MARCS per \citep{Gustafsson:1975p4658} & ARES & ABON2 via Spite 1967 (and improvements in the last 30yrs) & \citet{Trevisan:2011p6253} & 97 / 9 &  65\\ 
\citet{TucciMaia16} & Magellan Inamori Kyocera Echelle (MIKE) spectrograph (Bernstein et al. 2003) on the 6.5m Clay Magellan Telescope at Las Campanas Observatory & $>$65000 & 400 & 3200-10000 & MARCS per \citep{Gustafsson:1975p4658} & IRAF splot & MOOG14 per \citet{Sneden73} & differential analysis & 91 / 19 \\
\citet{Yan16} & FOCES echelle spectrograph on the 2.2 m telescope at Calar Alto Observatory & $>$ 40000 & $>$ 100 & 3700-9800  & MAFAGS & spectral fitting & spectral fitting & Cu = 4.25; Fe = 7.51 & 0 / 8  &  32\\ 
\citet{Zenoviene15} & Fibre-fed Echelle Spectrograph (FIES) on the Nordic Optical 2.5 m telescope & 68000 & $>$ 100 & 3680-7270  & MARCS per \citep{Gustafsson:1975p4658} & spectral fitting (Uppsala EQWIDTH) & spectral fitting (BSYN) & differential analysis & N/A &  44\\ 
\citet{Zhao16} & 1.52m telescope at ESO using FEROS & 48000 & 100 & 3560-9200  & MARCS per \citep{Gustafsson:1975p4658} & ARES & ABON2 via Spite 1967 (and improvements in the last 30yrs) & \citet{Anders:1989p3165} & 97 / 9 &  35\\
\hline
\tablenotetext{*}{Telescope/spectrograph information and the
techniques for determining abundances as given by the recently added
literature sources (with more than 20 stars added) into {\it
Hypatia} 2.1. Please see \citet{Hinkel14, Hinkel16} for more
details. In the header, ``S/N" is signal-to-noise report by the literature source, ``$\lambda$ Range" is the wavelength coverage, ``Stellar Atmo" is the stellar atmospheric model, ``Eq. Width" is the package used to determine the equivalent width, ``CoG or SF" designates whether the group used a curve-of-growth or spectral fitting technique where the package is specified in the former case, the ``Solar Scale" is the solar normalization used by that group (differential analysis is cited where applicable), and ``Num. of Fe I/II lines" lists the number of Fe I and Fe II lines. }

\enddata
\end{deluxetable}
\clearpage
\end{landscape}

 \clearpage
\LongTables
\begin{deluxetable*}{p{1.4cm}p{7.0cm}}
  \tablecaption{\label{params} Parameters in CATSUP}\tablenotetext{*}{All values of 99.99 are null.}
  \tablehead{
    \colhead{Column Header} &
    \colhead{Description}  
  }
 \startdata
HIP &  Hipparcos name \\
HD & Henry-Draper catalog name \\
TYC &  TYCHO name \\
RAJ2000 &  Right ascension (epoch = 2000) \\
DEJ2000 &  Declination (epoch = 2000) \\
X &   geocentric x-coordinate from the Sun, in pc\\
Y &   geocentric y-coordinate from the Sun, in pc \\
Z &   geocentric z-coordinate from the Sun, in pc\\
Uvel  & velocity (km/s) positive toward the Galactic anticenter (radial) \\
Vvel & velocity (km/s) positive in the direction of Galactic rotation \\
Wvel & velocity (km/s) positive toward the North Galactic Pole \\
Dist &   distance in pc (from Gaia) \\
Teff &   effective temperature of the star, in K \\
TeffSrc &   effective temperature reference source , namely PASTEL \citep[][and references therein]{Soubiran16}, ExoCat \citep[][and V-K data, see paper]{Gray03, Gray06, Valenti:2005p1491,Takeda:2007p1531} \\
logg &   surface gravity of the star \\
loggSrc &  surface gravity reference source , namely PASTEL \citep[][and references therein]{Soubiran16}, ExoCat \citep[][and V-K data, see paper]{Gray03, Gray06, Valenti:2005p1491,Takeda:2007p1531} \\
Disk &  likely origin within the disk (thin, thick, N/A) based on kinematics\\
Planet & flag (0, 1) as to whether a planet is known to orbit the star at the time of this publication, based on the NASA Exoplanet Archive \\
Bmag &  B magnitude  \\
Vmag &   V magnitude \\
BV &  B-V color   \\
Single &  a true single star (=1)  \\
Component &  if known to be a member of a multiple, to which component the HIP number is referring (A, B, etc). If the HIP number includes more than one star, either a known or suspected unresolved/very faint companion, noted by a ``+" symbol \\
HIP2 &  if more than one star in the system has its own HIP number, the other associated HIP numbers  \\
SpType &   spectral type \\
SpecSrc &  spectral type reference source, where the ADS suffix is provided in all cases. An asterisk (*) at the end of the reference indicates that the Houk spectral type was adjusted to modern MK system (using \citealt{Gray03,Gray06} spectral types) following \citet{Pecaut16}.  \\
logRHK &  the average value of logR'$_{HK}$ Ca II HK emission indices derived from literature sources \\
logRHKsources   & number of sources compiled for the logR'$_{HK}$ Ca II HK emission indices \\
FeH &  [Fe/H] abundance in dex \\
spFeH & spread in [Fe/H] abundance  \\
CH &   [C/H] abundance in dex\\
spCH & spread in [C/H] abundance  \\
OH &   [O/H] abundance in dex\\
spOH &  spread in [O/H] abundance \\
NaH &  [Na/H] abundance in dex \\
spNaH &  spread in [Na/H] abundance \\
MgH &   [Mg/H] abundance in dex\\
spMgH &  spread in [Mg/H] abundance \\
AlH &   [Al/H] abundance in dex\\
spAlH &  spread in [Al/H] abundance \\
SiH &   [Si/H] abundance in dex\\
spSiH &  spread in [Si/H] abundance \\
CaH &  [Ca/H] abundance in dex \\
spCaH & spread in [Ca/H] abundance  \\
TiH &  [Ti/H] abundance in dex \\
spTiH &  spread in [Ti/H] abundance \\
VH &  [V/H] abundance in dex \\
spVH &  spread in [V/H] abundance \\
CrH &  [Cr/H] abundance in dex\\
spCrH &  spread in [Cr/H] abundance \\
MnH &  [Mn/H] abundance in dex\\
spMnH &  spread in [Mn/H] abundance \\
CoH &  [Co/H] abundance in dex \\
spCoH &  spread in [Co/H] abundance \\
NiH &  [Ni/H] abundance in dex \\
spNiH &  spread in [Ni/H] abundance \\
FUVmag &  FUV magnitude (AB mag) \\
FUVmagerr &  error in FUV magnitude (AB mag)  \\
NUVmag &   NUV magnitude  (AB mag)\\
NUVmagerr & error in NUV magnitude  (AB mag) \\
l\_limit &   lower limit flag for FUV flux  \\
FUVflux & FUV flux ($\mu$Jy)  \\
FUVfluxerr & error in FUV flux ($\mu$Jy)  \\
u\_limit &  upper limit flag for NUV flux  \\
NUVflux &  NUV flux ($\mu$Jy)  \\
NUVfluxerr &  error in NUV flux ($\mu$Jy) \\
logRx    & $\log R_x$, where $R_x = L_x/L_{bol}$, X-ray to bolometric luminosity ratio or the fractional X-ray luminosity \\
logLLsun & $\log (L/L_{\odot})$, bolometric luminosity in Suns**\\
Lbol     & $L_{bol}$, bolometric luminosity\\
fROSAT   & log10 of X-ray flux from ROSAT (erg\,s$^{-1}$\,cm$^{-2}$)\\
fsROSAT  & log10 of X-ray surface flux from ROSAT (erg\,s$^{-1}$\,cm$^{-2}$)\\
LROSAT   & log10 of X-ray luminosity from ROSAT (erg\,s$^{-1}$) \\
RxROSAT  & log10 of $R_x (= L_x/L_{bol})$ from ROSAT\\
fXMM     & log10 of X-ray flux from XMM (erg\,s$^{-1}$\,cm$^{-2}$)\\
fsXMM    & log10 of X-ray surface flux from XMM (erg\,s$^{-1}$\,cm$^{-2}$)\\
LXMM     & log10 of X-ray luminosity from XMM (erg\,s$^{-1}$)\\
RxXMM    & log10 of $R_x (= L_x/L_{bol})$ XMM\\
fChan    & log10 of X-ray flux from Chandra (erg\,s$^{-1}$\,cm$^{-2}$)\\
fsChan   & log10 of X-ray surface flux from Chandra (erg\,s$^{-1}$\,cm$^{-2}$)\\
LChan    & log10 of X-ray luminosity from Chandra (erg\,s$^{-1}$)\\
RxChan   & log10 of $R_x (= L_x/L_{bol})$ from Chandra\\
\tablenotetext{**}{in units of the IAU nominal solar luminosity: L$_{Sun}$ =
  3.828\,$\times$\,10$^{26}$ W  \citep{Prsa16}}
 \enddata
 \end{deluxetable*}

\clearpage


\begin{thebibliography}{}
\expandafter\ifx\csname natexlab\endcsname\relax\def\natexlab#1{#1}\fi

\bibitem[{{Adibekyan} {et~al.}(2016){Adibekyan}, {Delgado-Mena}, {Figueira},
  {Sousa}, {Santos}, {Gonz{\'a}lez Hern{\'a}ndez}, {Minchev}, {Faria},
  {Israelian}, {Harutyunyan}, {Su{\'a}rez-Andr{\'e}s}, \&
  {Hakobyan}}]{Adibekyan16}
{Adibekyan}, V., {Delgado-Mena}, E., {Figueira}, P., {et~al.} 2016, \aap, 592,
  A87

\bibitem[{Anders \& Grevesse(1989)}]{Anders:1989p3165}
Anders, E., \& Grevesse, N. 1989, Geochimica et Cosmochimica Acta, 53, 197

\bibitem[{{Anderson} \& {Francis}(2012)}]{Anderson12}
{Anderson}, E., \& {Francis}, C. 2012, Astronomy Letters, 38, 331

\bibitem[{{Arenou} {et~al.}(2017){Arenou}, {Luri}, {Babusiaux}, {Fabricius},
  {Helmi}, {Robin}, {Vallenari}, {Blanco-Cuaresma}, {Cantat-Gaudin},
  {Findeisen}, {Reyl{\'e}}, {Ruiz-Dern}, {Sordo}, {Turon}, {Walton}, {Shih},
  {Antiche}, {Barache}, {Barros}, {Breddels}, {Carrasco}, {Costigan},
  {Diakit{\'e}}, {Eyer}, {Figueras}, {Galluccio}, {Heu}, {Jordi},
  {Krone-Martins}, {Lallement}, {Lambert}, {Leclerc}, {Marrese}, {Moitinho},
  {Mor}, {Romero-G{\'o}mez}, {Sartoretti}, {Soria}, {Soubiran}, {Souchay},
  {Veljanoski}, {Ziaeepour}, {Giuffrida}, {Pancino}, \& {Bragaglia}}]{Arenou17}
{Arenou}, F., {Luri}, X., {Babusiaux}, C., {et~al.} 2017, \aap, 599, A50

\bibitem[{{Arney} {et~al.}(2016){Arney}, {Domagal-Goldman}, {Meadows}, {Wolf},
  {Schwieterman}, {Charnay}, {Claire}, {H{\'e}brard}, \& {Trainer}}]{Arney16}
{Arney}, G., {Domagal-Goldman}, S.~D., {Meadows}, V.~S., {et~al.} 2016,
  Astrobiology, 16, 873

\bibitem[{Asplund {et~al.}(2009)Asplund, Grevesse, Sauval, \&
  Scott}]{Asplund:2009p3251}
Asplund, M., Grevesse, N., Sauval, A.~J., \& Scott, P. 2009, ARA\&A, 47, 481

\bibitem[{{Astraatmadja} \& {Bailer-Jones}(2016)}]{Astraatmadja16}
{Astraatmadja}, T.~L., \& {Bailer-Jones}, C.~A.~L. 2016, \apj, 833, 119

\bibitem[{{Batalha} {et~al.}(2013){Batalha}, {Rowe}, {Bryson}, {Barclay},
  {Burke}, {Caldwell}, {Christiansen}, {Mullally}, {Thompson}, {Brown},
  {Dupree}, {Fabrycky}, {Ford}, {Fortney}, {Gilliland}, {Isaacson}, {Latham},
  {Marcy}, {Quinn}, {Ragozzine}, {Shporer}, {Borucki}, {Ciardi}, {Gautier},
  {Haas}, {Jenkins}, {Koch}, {Lissauer}, {Rapin}, {Basri}, {Boss}, {Buchhave},
  {Carter}, {Charbonneau}, {Christensen-Dalsgaard}, {Clarke}, {Cochran},
  {Demory}, {Desert}, {Devore}, {Doyle}, {Esquerdo}, {Everett}, {Fressin},
  {Geary}, {Girouard}, {Gould}, {Hall}, {Holman}, {Howard}, {Howell},
  {Ibrahim}, {Kinemuchi}, {Kjeldsen}, {Klaus}, {Li}, {Lucas}, {Meibom},
  {Morris}, {Pr{\v s}a}, {Quintana}, {Sanderfer}, {Sasselov}, {Seader},
  {Smith}, {Steffen}, {Still}, {Stumpe}, {Tarter}, {Tenenbaum}, {Torres},
  {Twicken}, {Uddin}, {Van Cleve}, {Walkowicz}, \& {Welsh}}]{Batalha13}
{Batalha}, N.~M., {Rowe}, J.~F., {Bryson}, S.~T., {et~al.} 2013, \apjs, 204, 24

\bibitem[{Battistini \& Bensby(2016)}]{Battistini16}
Battistini, C., \& Bensby, T. 2016, \aap, 586, A49

\bibitem[{Baumann {et~al.}(2010)Baumann, Ram{\'{\i}}rez, Mel{\'{e}}ndez,
  Asplund, \& Lind}]{Baumann10}
Baumann, P., Ram{\'{\i}}rez, I., Mel{\'{e}}ndez, J., Asplund, M., \& Lind, K.
  2010, \aap, 519, A87

\bibitem[{Bensby {et~al.}(2003)Bensby, Feltzing, \& Lundstr{\"o}m}]{Bensby03}
Bensby, T., Feltzing, S., \& Lundstr{\"o}m, I. 2003, A{\&}A, 410, 527

\bibitem[{{Blanco-Cuaresma} {et~al.}(2014){Blanco-Cuaresma}, {Soubiran},
  {Heiter}, \& {Jofr{\'e}}}]{Blanco14}
{Blanco-Cuaresma}, S., {Soubiran}, C., {Heiter}, U., \& {Jofr{\'e}}, P. 2014,
  \aap, 569, A111

\bibitem[{{Boller} {et~al.}(2016){Boller}, {Freyberg}, {Tr{\"u}mper}, {Haberl},
  {Voges}, \& {Nandra}}]{Boller16}
{Boller}, T., {Freyberg}, M.~J., {Tr{\"u}mper}, J., {et~al.} 2016, \aap, 588,
  A103

\bibitem[{{Brewer} {et~al.}(2016){Brewer}, {Fischer}, {Valenti}, \&
  {Piskunov}}]{Brewer16}
{Brewer}, J.~M., {Fischer}, D.~A., {Valenti}, J.~A., \& {Piskunov}, N. 2016,
  \apjs, 225, 32

\bibitem[{{Burger} {et~al.}(2013){Burger}, {Stassun}, {Pepper}, {Siverd},
  {Paegert}, {De Lee}, \& {Robinson}}]{Burger13}
{Burger}, D., {Stassun}, K.~G., {Pepper}, J., {et~al.} 2013, Astronomy and
  Computing, 2, 40

\bibitem[{Chen {et~al.}(2001)Chen, Nissen, Benoni, \& Zhao}]{Chen01}
Chen, Y.~Q., Nissen, P.~E., Benoni, T., \& Zhao, G. 2001, \aap, 371, 943

\bibitem[{{da Silva} {et~al.}(2015){da Silva}, {Milone}, \&
  {Rocha-Pinto}}]{daSilva15}
{da Silva}, R., {Milone}, A.~d.~C., \& {Rocha-Pinto}, H.~J. 2015, \aap, 580,
  A24

\bibitem[{{Delgado Mena} {et~al.}(2014){Delgado Mena}, {Israelian},
  {Gonz{\'a}lez Hern{\'a}ndez}, {Sousa}, {Mortier}, {Santos}, {Adibekyan},
  {Fernandes}, {Rebolo}, {Udry}, \& {Mayor}}]{DelgadoMena14}
{Delgado Mena}, E., {Israelian}, G., {Gonz{\'a}lez Hern{\'a}ndez}, J.~I.,
  {et~al.} 2014, \aap, 562, A92

\bibitem[{{Delgado Mena} {et~al.}(2015){Delgado Mena}, {Bertr{\'a}n de Lis},
  {Adibekyan}, {Sousa}, {Figueira}, {Mortier}, {Gonz{\'a}lez Hern{\'a}ndez},
  {Tsantaki}, {Israelian}, \& {Santos}}]{DelgadoMena15}
{Delgado Mena}, E., {Bertr{\'a}n de Lis}, S., {Adibekyan}, V.~Z., {et~al.}
  2015, \aap, 576, A69

\bibitem[{{Dressing} \& {Charbonneau}(2013)}]{Dressing13}
{Dressing}, C.~D., \& {Charbonneau}, D. 2013, \apj, 767, 95

\bibitem[{{Evans} {et~al.}(2010){Evans}, {Primini}, {Glotfelty}, {Anderson},
  {Bonaventura}, {Chen}, {Davis}, {Doe}, {Evans}, {Fabbiano}, {Galle}, {Gibbs},
  {Grier}, {Hain}, {Hall}, {Harbo}, {(Helen He}, {Houck}, {Karovska},
  {Kashyap}, {Lauer}, {McCollough}, {McDowell}, {Miller}, {Mitschang},
  {Morgan}, {Mossman}, {Nichols}, {Nowak}, {Plummer}, {Refsdal}, {Rots},
  {Siemiginowska}, {Sundheim}, {Tibbetts}, {Van Stone}, {Winkelman}, \&
  {Zografou}}]{Evans10}
{Evans}, I.~N., {Primini}, F.~A., {Glotfelty}, K.~J., {et~al.} 2010, \apjs,
  189, 37

\bibitem[{{Evans} {et~al.}(2014){Evans}, {Primini}, {Glotfelty}, {Anderson},
  {Bonaventura}, {Chen}, {Davis}, {Doe}, {Evans}, {Fabbiano}, {Galle}, {Gibbs},
  {Grier}, {Hain}, {Hall}, {Harbo}, {He}, {Houck}, {Karovska}, {Kashyap},
  {Lauer}, {McCollough}, {McDowell}, {Miller}, {Mitschang}, {Morgan},
  {Mossman}, {Nichols}, {Nowak}, {Plummer}, {Refsdal}, {Rots}, {Siemiginowska},
  {Sundheim}, {Tibbetts}, {van Stone}, {Winkelman}, \& {Zografou}}]{Evans12}
{Evans}, I.~N., {Primini}, F.~A., {Glotfelty}, C.~S., {et~al.} 2014, VizieR
  Online Data Catalog, 9045

\bibitem[{{Fabricius} {et~al.}(2002){Fabricius}, {H{\o}g}, {Makarov}, {Mason},
  {Wycoff}, \& {Urban}}]{Fabricius02}
{Fabricius}, C., {H{\o}g}, E., {Makarov}, V.~V., {et~al.} 2002, \aap, 384, 180

\bibitem[{{Findeisen} {et~al.}(2011){Findeisen}, {Hillenbrand}, \&
  {Soderblom}}]{Findeisen11}
{Findeisen}, K., {Hillenbrand}, L., \& {Soderblom}, D. 2011, \aj, 142, 23

\bibitem[{{Fleming} {et~al.}(1995){Fleming}, {Molendi}, {Maccacaro}, \&
  {Wolter}}]{Fleming95}
{Fleming}, T.~A., {Molendi}, S., {Maccacaro}, T., \& {Wolter}, A. 1995, \apjs,
  99, 701

\bibitem[{{Gaia Collaboration} {et~al.}(2016){Gaia Collaboration}, {Brown},
  {Vallenari}, {Prusti}, {de Bruijne}, {Mignard}, {Drimmel}, {Babusiaux},
  {Bailer-Jones}, {Bastian}, \& et~al.}]{Gaia16}
{Gaia Collaboration}, {Brown}, A.~G.~A., {Vallenari}, A., {et~al.} 2016, \aap,
  595, A2

\bibitem[{{Garmire} {et~al.}(2003){Garmire}, {Bautz}, {Ford}, {Nousek}, \&
  {Ricker}}]{Garmire03}
{Garmire}, G.~P., {Bautz}, M.~W., {Ford}, P.~G., {Nousek}, J.~A., \& {Ricker},
  Jr., G.~R. 2003, in \procspie, Vol. 4851, X-Ray and Gamma-Ray Telescopes and
  Instruments for Astronomy., ed. J.~E. {Truemper} \& H.~D. {Tananbaum}, 28--44

\bibitem[{Ghezzi {et~al.}(2010)Ghezzi, Cunha, Smith, \& de~la Reza}]{Ghezzi10b}
Ghezzi, L., Cunha, K., Smith, V.~V., \& de~la Reza, R. 2010, The Astrophysical
  Journal, 724, 154

\bibitem[{{Gomes da Silva} {et~al.}(2014){Gomes da Silva}, {Santos}, {Boisse},
  {Dumusque}, \& {Lovis}}]{daSilva14}
{Gomes da Silva}, J., {Santos}, N.~C., {Boisse}, I., {Dumusque}, X., \&
  {Lovis}, C. 2014, \aap, 566, A66

\bibitem[{Gonzalez(2014)}]{Gonzalez14}
Gonzalez, G. 2014, Monthly Notices of the Royal Astronomical Society, 441, 1201

\bibitem[{Gonzalez(2015)}]{Gonzalez15}
---. 2015, Monthly Notices of the Royal Astronomical Society, 446, 1020

\bibitem[{Gonzalez {et~al.}(2010)Gonzalez, Carlson, \& Tobin}]{Gonzalez10a}
Gonzalez, G., Carlson, M.~K., \& Tobin, R.~W. 2010, Monthly Notices of the
  Royal Astronomical Society, 403, 1368

\bibitem[{{Gonzalez} {et~al.}(2010){Gonzalez}, {Carlson}, \&
  {Tobin}}]{Gonzalez10b}
{Gonzalez}, G., {Carlson}, M.~K., \& {Tobin}, R.~W. 2010, \mnras, 407, 314

\bibitem[{{Gonzalez} {et~al.}(2001){Gonzalez}, {Laws}, {Tyagi}, \&
  {Reddy}}]{Gonzalez01}
{Gonzalez}, G., {Laws}, C., {Tyagi}, S., \& {Reddy}, B.~E. 2001, \aj, 121, 432

\bibitem[{{Gray}(1997)}]{Gray97}
{Gray}, D.~F. 1997, \nat, 385, 795

\bibitem[{{Gray} {et~al.}(2006){Gray}, {Corbally}, {Garrison}, {McFadden},
  {Bubar}, {McGahee}, {O'Donoghue}, \& {Knox}}]{Gray06}
{Gray}, R.~O., {Corbally}, C.~J., {Garrison}, R.~F., {et~al.} 2006, \aj, 132,
  161

\bibitem[{{Gray} {et~al.}(2003){Gray}, {Corbally}, {Garrison}, {McFadden}, \&
  {Robinson}}]{Gray03}
{Gray}, R.~O., {Corbally}, C.~J., {Garrison}, R.~F., {McFadden}, M.~T., \&
  {Robinson}, P.~E. 2003, \aj, 126, 2048

\bibitem[{Gustafsson {et~al.}(1975)Gustafsson, Bell, Eriksson, \&
  Nordlund}]{Gustafsson:1975p4658}
Gustafsson, B., Bell, R.~A., Eriksson, K., \& Nordlund, A. 1975, A{\&}A, 42,
  407

\bibitem[{{Hartmann} {et~al.}(1984){Hartmann}, {Soderblom}, {Noyes}, {Burnham},
  \& {Vaughan}}]{Hartmann84}
{Hartmann}, L., {Soderblom}, D.~R., {Noyes}, R.~W., {Burnham}, N., \&
  {Vaughan}, A.~H. 1984, \apj, 276, 254

\bibitem[{{Henry} {et~al.}(1996){Henry}, {Soderblom}, {Donahue}, \&
  {Baliunas}}]{Henry96}
{Henry}, T.~J., {Soderblom}, D.~R., {Donahue}, R.~A., \& {Baliunas}, S.~L.
  1996, \aj, 111, doi:10.1086/117796

\bibitem[{{Henry} {et~al.}(2002){Henry}, {Walkowicz}, {Barto}, \&
  {Golimowski}}]{Henry02}
{Henry}, T.~J., {Walkowicz}, L.~M., {Barto}, T.~C., \& {Golimowski}, D.~A.
  2002, \aj, 123, 2002

\bibitem[{{Hinkel} {et~al.}(2014){Hinkel}, {Timmes}, {Young}, {Pagano}, \&
  {Turnbull}}]{Hinkel14}
{Hinkel}, N.~R., {Timmes}, F.~X., {Young}, P.~A., {Pagano}, M.~D., \&
  {Turnbull}, M.~C. 2014, \aj, 148, 54

\bibitem[{{Hinkel} {et~al.}(2016){Hinkel}, {Young}, {Pagano}, {Desch}, {Anbar},
  {Adibekyan}, {Blanco-Cuaresma}, {Carlberg}, {Delgado Mena}, {Liu},
  {Nordlander}, {Sousa}, {Korn}, {Gruyters}, {Heiter}, {Jofr{\'e}}, {Santos},
  \& {Soubiran}}]{Hinkel16}
{Hinkel}, N.~R., {Young}, P.~A., {Pagano}, M.~D., {et~al.} 2016, \apjs, 226, 4

\bibitem[{{Houk} \& {Cowley}(1975)}]{Houk75}
{Houk}, N., \& {Cowley}, A.~P. 1975, {University of Michigan Catalogue of
  two-dimensional spectral types for the HD stars. Volume I. Declinations -90
  to -53} (Dept. of Astronomy, University of Michigan)

\bibitem[{Israelian {et~al.}(2004)Israelian, Santos, Mayor, \&
  Rebolo}]{Israelian04}
Israelian, G., Santos, N.~C., Mayor, M., \& Rebolo, R. 2004, Astronomy and
  Astrophysics, 414, 601

\bibitem[{Israelian {et~al.}(2009)Israelian, Mena, Santos, Sousa, Mayor, Udry,
  Cerde{\~{n}}a, Rebolo, \& Randich}]{Israelian09}
Israelian, G., Mena, E.~D., Santos, N.~C., {et~al.} 2009, Nature, 462, 189

\bibitem[{{Jenkins} {et~al.}(2008){Jenkins}, {Jones}, {Pavlenko}, {Pinfield},
  {Barnes}, \& {Lyubchik}}]{Jenkins08}
{Jenkins}, J.~S., {Jones}, H.~R.~A., {Pavlenko}, Y., {et~al.} 2008, \aap, 485,
  571

\bibitem[{{Jenkins} {et~al.}(2006){Jenkins}, {Jones}, {Tinney}, {Butler},
  {McCarthy}, {Marcy}, {Pinfield}, {Carter}, \& {Penny}}]{Jenkins06}
{Jenkins}, J.~S., {Jones}, H.~R.~A., {Tinney}, C.~G., {et~al.} 2006, \mnras,
  372, 163

\bibitem[{{Jenkins} {et~al.}(2011){Jenkins}, {Murgas}, {Rojo}, {Jones},
  {Day-Jones}, {Jones}, {Clarke}, {Ruiz}, \& {Pinfield}}]{Jenkins11}
{Jenkins}, J.~S., {Murgas}, F., {Rojo}, P., {et~al.} 2011, \aap, 531, A8

\bibitem[{{Johnson} {et~al.}(2016){Johnson}, {Endl}, {Cochran}, {Meschiari},
  {Robertson}, {MacQueen}, {Brugamyer}, {Caldwell}, {Hatzes}, {Ram{\'{\i}}rez},
  \& {Wittenmyer}}]{Johnson16}
{Johnson}, M.~C., {Endl}, M., {Cochran}, W.~D., {et~al.} 2016, \apj, 821, 74

\bibitem[{{Johnstone} \& {G{\"u}del}(2015)}]{Johnstone15}
{Johnstone}, C.~P., \& {G{\"u}del}, M. 2015, \aap, 578, A129

\bibitem[{{Keenan} \& {McNeil}(1989)}]{Keenan89}
{Keenan}, P.~C., \& {McNeil}, R.~C. 1989, \apjs, 71, 245

\bibitem[{{Kurucz}(1993)}]{Kurucz1993}
{Kurucz}, R.~L. 1993, SYNTHE spectrum synthesis programs and line data
  (Cambridge, MA: Smithsonian Astrophysical Observatory, |c1993)

\bibitem[{{Lambert} {et~al.}(1991){Lambert}, {Heath}, \&
  {Edvardsson}}]{Lambert91}
{Lambert}, D.~L., {Heath}, J.~E., \& {Edvardsson}, B. 1991, \mnras, 253, 610

\bibitem[{Lambert \& Reddy(2004)}]{Lambert04}
Lambert, D.~L., \& Reddy, B.~E. 2004, Monthly Notices of the Royal Astronomical
  Society, 349, 757

\bibitem[{{Linsky} {et~al.}(2000){Linsky}, {Redfield}, {Wood}, \&
  {Piskunov}}]{Linsky00}
{Linsky}, J.~L., {Redfield}, S., {Wood}, B.~E., \& {Piskunov}, N. 2000, \apj,
  528, 756

\bibitem[{Liu {et~al.}(2014)Liu, Tan, Wang, Zhao, Sato, Takeda, \& Li}]{Liu14}
Liu, Y.~J., Tan, K.~F., Wang, L., {et~al.} 2014, The Astrophysical Journal,
  785, 94

\bibitem[{Lodders {et~al.}(2009)Lodders, Plame, \& Gail}]{Lodders:2009p3091}
Lodders, K., Plame, H., \& Gail, H.-P. 2009, Abundances of the Elements in the
  Solar System, Vol.~4B (Berlin, Heidelberg, New York: Springer-Verlag), 44

\bibitem[{L{\'{o}}pez-Valdivia {et~al.}(2017)L{\'{o}}pez-Valdivia, Bertone, \&
  Ch{\'{a}}vez}]{LopezValdivia17}
L{\'{o}}pez-Valdivia, R., Bertone, E., \& Ch{\'{a}}vez, M. 2017, Monthly
  Notices of the Royal Astronomical Society, stx249

\bibitem[{Luck(2015)}]{Luck15}
Luck, R.~E. 2015, The Astronomical Journal, 150, 88

\bibitem[{Luck(2017)}]{Luck17}
---. 2017, The Astronomical Journal, 153, 21

\bibitem[{{Luck} \& {Heiter}(2006)}]{Luck06}
{Luck}, R.~E., \& {Heiter}, U. 2006, \aj, 131, 3069

\bibitem[{{Luck} \& {Heiter}(2007)}]{Luck07}
---. 2007, \aj, 133, 2464

\bibitem[{{Luger} \& {Barnes}(2015)}]{luge15}
{Luger}, R., \& {Barnes}, R. 2015, Astrobiology, 15, 119

\bibitem[{{Mahdi} {et~al.}(2016){Mahdi}, {Soubiran}, {Blanco-Cuaresma}, \&
  {Chemin}}]{Mahdi16}
{Mahdi}, D., {Soubiran}, C., {Blanco-Cuaresma}, S., \& {Chemin}, L. 2016, \aap,
  587, A131

\bibitem[{Maldonado {et~al.}(2015)Maldonado, Eiroa, Villaver, Montesinos, \&
  Mora}]{Maldonado15}
Maldonado, J., Eiroa, C., Villaver, E., Montesinos, B., \& Mora, A. 2015,
  Astronomy {\&} Astrophysics, 579, A20

\bibitem[{{Maldonado} \& {Villaver}(2016)}]{Maldonado16}
{Maldonado}, J., \& {Villaver}, E. 2016, \aap, 588, A98

\bibitem[{Mallik {et~al.}(2003)Mallik, Parthasarathy, \& Pati}]{Mallik03}
Mallik, S.~V., Parthasarathy, M., \& Pati, A.~K. 2003, Astronomy and
  Astrophysics, 409, 251

\bibitem[{{Mamajek} \& {Hillenbrand}(2008)}]{Mamajek08}
{Mamajek}, E.~E., \& {Hillenbrand}, L.~A. 2008, \apj, 687, 1264

\bibitem[{{Mason} {et~al.}(2001){Mason}, {Wycoff}, {Hartkopf}, {Douglass}, \&
  {Worley}}]{Mason01}
{Mason}, B.~D., {Wycoff}, G.~L., {Hartkopf}, W.~I., {Douglass}, G.~G., \&
  {Worley}, C.~E. 2001, \aj, 122, 3466

\bibitem[{{Mayor} {et~al.}(2011){Mayor}, {Marmier}, {Lovis}, {Udry},
  {S{\'e}gransan}, {Pepe}, {Benz}, {Bertaux}, {Bouchy}, {Dumusque}, {Lo Curto},
  {Mordasini}, {Queloz}, \& {Santos}}]{Mayor11}
{Mayor}, M., {Marmier}, M., {Lovis}, C., {et~al.} 2011, ArXiv e-prints,
  arXiv:1109.2497

\bibitem[{Mennesson {et~al.}(2016)Mennesson, Gaudi, Seager, Cahoy,
  Domagal-Goldman, Feinberg, Guyon, Kasdin, Marois, Mawet, Tamura, Mouillet,
  Prusti, Quirrenbach, Robinson, Rogers, Scowen, Somerville, Stapelfeldt,
  Stern, Still, Turnbull, Booth, Kiessling, Kuan, \& Warfield}]{Mennesson16}
Mennesson, B., Gaudi, S., Seager, S., {et~al.} 2016, in Space Telescopes and
  Instrumentation 2016: Optical, Infrared, and Millimeter Wave, ed. H.~A.
  MacEwen, G.~G. Fazio, M.~Lystrup, N.~Batalha, N.~Siegler, \& E.~C. Tong
  ({SPIE})

\bibitem[{{Middelkoop}(1982)}]{Middelkoop82}
{Middelkoop}, F. 1982, \aap, 107, 31

\bibitem[{{Miguel} \& {Kaltenegger}(2014)}]{migu14a}
{Miguel}, Y., \& {Kaltenegger}, L. 2014, \apj, 780, 166, 13 pp

\bibitem[{{Mishenina} {et~al.}(2016){Mishenina}, {Kovtyukh}, {Soubiran}, \&
  {Adibekyan}}]{Mishenina16}
{Mishenina}, T., {Kovtyukh}, V., {Soubiran}, C., \& {Adibekyan}, V.~Z. 2016,
  \mnras, 462, 1563

\bibitem[{{Mishenina} {et~al.}(2012){Mishenina}, {Soubiran}, {Kovtyukh},
  {Katsova}, \& {Livshits}}]{Mishenina12}
{Mishenina}, T.~V., {Soubiran}, C., {Kovtyukh}, V.~V., {Katsova}, M.~M., \&
  {Livshits}, M.~A. 2012, \aap, 547, A106

\bibitem[{{Morrissey} {et~al.}(2005){Morrissey}, {Schiminovich}, {Barlow},
  {Martin}, {Blakkolb}, {Conrow}, {Cooke}, {Erickson}, {Fanson}, {Friedman},
  {Grange}, {Jelinsky}, {Lee}, {Liu}, {Mazer}, {McLean}, {Milliard}, {Randall},
  {Schmitigal}, {Sen}, {Siegmund}, {Surber}, {Vaughan}, {Viton}, {Welsh},
  {Bianchi}, {Byun}, {Donas}, {Forster}, {Heckman}, {Lee}, {Madore}, {Malina},
  {Neff}, {Rich}, {Small}, {Szalay}, \& {Wyder}}]{morr05}
{Morrissey}, P., {Schiminovich}, D., {Barlow}, T.~A., {et~al.} 2005, \apjl,
  619, L7

\bibitem[{{Morrissey} {et~al.}(2007){Morrissey}, {Conrow}, {Barlow}, {Small},
  {Seibert}, {Wyder}, {Budav{\'a}ri}, {Arnouts}, {Friedman}, {Forster},
  {Martin}, {Neff}, {Schiminovich}, {Bianchi}, {Donas}, {Heckman}, {Lee},
  {Madore}, {Milliard}, {Rich}, {Szalay}, {Welsh}, \& {Yi}}]{morr07}
{Morrissey}, P., {Conrow}, T., {Barlow}, T.~A., {et~al.} 2007, \apjs, 173, 682

\bibitem[{{Mukai} \& {Shiokawa}(1993)}]{Mukai93}
{Mukai}, K., \& {Shiokawa}, K. 1993, \apj, 418, 863

\bibitem[{{Neuhaeuser} {et~al.}(1995){Neuhaeuser}, {Sterzik}, {Schmitt},
  {Wichmann}, \& {Krautter}}]{Neuhauser95}
{Neuhaeuser}, R., {Sterzik}, M.~F., {Schmitt}, J.~H.~M.~M., {Wichmann}, R., \&
  {Krautter}, J. 1995, \aap, 297, 391

\bibitem[{Nissen(2013)}]{Nissen13}
Nissen, P.~E. 2013, Astronomy {\&} Astrophysics, 552, A73

\bibitem[{{Nissen}(2016)}]{Nissen16}
{Nissen}, P.~E. 2016, \aap, 593, A65

\bibitem[{{Notsu} {et~al.}(2017){Notsu}, {Honda}, {Maehara}, {Notsu},
  {Namekata}, {Nogami}, \& {Shibata}}]{Notsu17}
{Notsu}, Y., {Honda}, S., {Maehara}, H., {et~al.} 2017, \pasj, 69, 12

\bibitem[{{Noyes} {et~al.}(1984){Noyes}, {Hartmann}, {Baliunas}, {Duncan}, \&
  {Vaughan}}]{Noyes84}
{Noyes}, R.~W., {Hartmann}, L.~W., {Baliunas}, S.~L., {Duncan}, D.~K., \&
  {Vaughan}, A.~H. 1984, \apj, 279, 763

\bibitem[{{Ochsenbein} {et~al.}(2000){Ochsenbein}, {Bauer}, \&
  {Marcout}}]{Ochsenbein00}
{Ochsenbein}, F., {Bauer}, P., \& {Marcout}, J. 2000, \aaps, 143, 23

\bibitem[{{Pagano} {et~al.}(2017){Pagano}, {Young}, \& {Challa}}]{Pagano17}
{Pagano}, M., {Young}, P.~A., \& {Challa}, P. 2017, \apj, in review

\bibitem[{{Pecaut} \& {Mamajek}(2016)}]{Pecaut16}
{Pecaut}, M.~J., \& {Mamajek}, E.~E. 2016, \mnras, 461, 794

\bibitem[{{Pr{\v s}a} {et~al.}(2016){Pr{\v s}a}, {Harmanec}, {Torres},
  {Mamajek}, {Asplund}, {Capitaine}, {Christensen-Dalsgaard}, {Depagne},
  {Haberreiter}, {Hekker}, {Hilton}, {Kopp}, {Kostov}, {Kurtz}, {Laskar},
  {Mason}, {Milone}, {Montgomery}, {Richards}, {Schmutz}, {Schou}, \&
  {Stewart}}]{Prsa16}
{Pr{\v s}a}, A., {Harmanec}, P., {Torres}, G., {et~al.} 2016, \aj, 152, 41

\bibitem[{{Ram{\'{\i}}rez} {et~al.}(2013){Ram{\'{\i}}rez}, {Allende Prieto}, \&
  {Lambert}}]{Ramirez13}
{Ram{\'{\i}}rez}, I., {Allende Prieto}, C., \& {Lambert}, D.~L. 2013, \apj,
  764, 78

\bibitem[{Ram{\'{\i}}rez {et~al.}(2012)Ram{\'{\i}}rez, Fish, Lambert, \&
  Prieto}]{Ramirez12a}
Ram{\'{\i}}rez, I., Fish, J.~R., Lambert, D.~L., \& Prieto, C.~A. 2012, The
  Astrophysical Journal, 756, 46

\bibitem[{Reddy {et~al.}(2003)Reddy, Tomkin, Lambert, \&
  Prieto}]{Reddy:2003p1354}
Reddy, B., Tomkin, J., Lambert, D., \& Prieto, C.~A. 2003, MNRAS, 340, 304

\bibitem[{{Reetz}(1991)}]{Reetz91}
{Reetz}, J.~K. 1991, PhD thesis, Ludwig-Maximilians-Universit{\"a}t M{\"u}nchen

\bibitem[{{Reis} {et~al.}(2011){Reis}, {Corradi}, {de Avillez}, \&
  {Santos}}]{Reis11}
{Reis}, W., {Corradi}, W., {de Avillez}, M.~A., \& {Santos}, F.~P. 2011, \apj,
  734, 8

\bibitem[{{Rosat}(2000)}]{ROSAT00}
{Rosat}, C. 2000, VizieR Online Data Catalog, 9030

\bibitem[{{Rosen} {et~al.}(2016){Rosen}, {Webb}, {Watson}, {Ballet}, {Barret},
  {Braito}, {Carrera}, {Ceballos}, {Coriat}, {Della Ceca}, {Denkinson},
  {Esquej}, {Farrell}, {Freyberg}, {Gris{\'e}}, {Guillout}, {Heil},
  {Koliopanos}, {Law-Green}, {Lamer}, {Lin}, {Martino}, {Michel}, {Motch},
  {Nebot Gomez-Moran}, {Page}, {Page}, {Page}, {Pakull}, {Pye}, {Read},
  {Rodriguez}, {Sakano}, {Saxton}, {Schwope}, {Scott}, {Sturm}, {Traulsen},
  {Yershov}, \& {Zolotukhin}}]{Rosen16}
{Rosen}, S.~R., {Webb}, N.~A., {Watson}, M.~G., {et~al.} 2016, \aap, 590, A1

\bibitem[{{Rugheimer} {et~al.}(2015){Rugheimer}, {Kaltenegger}, {Segura},
  {Linsky}, \& {Mohanty}}]{rugh15}
{Rugheimer}, S., {Kaltenegger}, L., {Segura}, A., {Linsky}, J., \& {Mohanty},
  S. 2015, \apj, 809, 57

\bibitem[{Seager {et~al.}(2015)Seager, Turnbull, Sparks, Thomson, Shaklan,
  Roberge, Kuchner, Kasdin, Domagal-Goldman, Cash, Warfield, Lisman, Scharf,
  Webb, Trabert, Martin, Cady, \& Heneghan}]{Seager15}
Seager, S., Turnbull, M., Sparks, W., {et~al.} 2015, in Techniques and
  Instrumentation for Detection of Exoplanets {VII}, ed. S.~Shaklan ({SPIE})

\bibitem[{{Shkolnik}(2013)}]{Shkolnik13}
{Shkolnik}, E.~L. 2013, \apj, 766, 9

\bibitem[{{Showman} \& {Malhotra}(1999)}]{Showman99}
{Showman}, A.~P., \& {Malhotra}, R. 1999, Science, 296, 77

\bibitem[{{Smith}(2011)}]{Smith11}
{Smith}, G. 2011, The Observatory, 131, 1

\bibitem[{{Smith} \& {Redenbaugh}(2010)}]{Smith10}
{Smith}, G.~H., \& {Redenbaugh}, A.~K. 2010, \pasp, 122, 1303

\bibitem[{{Sneden}(1973)}]{Sneden73}
{Sneden}, C.~A. 1973, PhD thesis, The University of Texas - Austin

\bibitem[{{Soderblom}(1985)}]{Soderblom85}
{Soderblom}, D.~R. 1985, \aj, 90, 2103

\bibitem[{{Soderblom}(1990)}]{Soderblom90}
---. 1990, \aj, 100, 204

\bibitem[{{Soderblom} {et~al.}(1991){Soderblom}, {Duncan}, \&
  {Johnson}}]{Soderblom91}
{Soderblom}, D.~R., {Duncan}, D.~K., \& {Johnson}, D.~R.~H. 1991, \apj, 375,
  722

\bibitem[{{Soubiran} {et~al.}(2016){Soubiran}, {Le Campion}, {Brouillet}, \&
  {Chemin}}]{Soubiran16}
{Soubiran}, C., {Le Campion}, J.-F., {Brouillet}, N., \& {Chemin}, L. 2016,
  \aap, 591, A118

\bibitem[{{Spergel} {et~al.}(2013){Spergel}, {Gehrels}, {Breckinridge},
  {Donahue}, {Dressler}, {Gaudi}, {Greene}, {Guyon}, {Hirata}, {Kalirai},
  {Kasdin}, {Moos}, {Perlmutter}, {Postman}, {Rauscher}, {Rhodes}, {Wang},
  {Weinberg}, {Centrella}, {Traub}, {Baltay}, {Colbert}, {Bennett},
  {Kiessling}, {Macintosh}, {Merten}, {Mortonson}, {Penny}, {Rozo},
  {Savransky}, {Stapelfeldt}, {Zu}, {Baker}, {Cheng}, {Content}, {Dooley},
  {Foote}, {Goullioud}, {Grady}, {Jackson}, {Kruk}, {Levine}, {Melton},
  {Peddie}, {Ruffa}, \& {Shaklan}}]{Spergel13}
{Spergel}, D., {Gehrels}, N., {Breckinridge}, J., {et~al.} 2013, ArXiv
  e-prints, arXiv:1305.5422

\bibitem[{{Spergel} {et~al.}(2015){Spergel}, {Gehrels}, {Baltay}, {Bennett},
  {Breckinridge}, {Donahue}, {Dressler}, {Gaudi}, {Greene}, {Guyon}, {Hirata},
  {Kalirai}, {Kasdin}, {Macintosh}, {Moos}, {Perlmutter}, {Postman},
  {Rauscher}, {Rhodes}, {Wang}, {Weinberg}, {Benford}, {Hudson}, {Jeong},
  {Mellier}, {Traub}, {Yamada}, {Capak}, {Colbert}, {Masters}, {Penny},
  {Savransky}, {Stern}, {Zimmerman}, {Barry}, {Bartusek}, {Carpenter}, {Cheng},
  {Content}, {Dekens}, {Demers}, {Grady}, {Jackson}, {Kuan}, {Kruk}, {Melton},
  {Nemati}, {Parvin}, {Poberezhskiy}, {Peddie}, {Ruffa}, {Wallace}, {Whipple},
  {Wollack}, \& {Zhao}}]{Spergel15}
{Spergel}, D., {Gehrels}, N., {Baltay}, C., {et~al.} 2015, ArXiv e-prints,
  arXiv:1503.03757

\bibitem[{{Stassun} {et~al.}(2017){Stassun}, {Oelkers}, {Pepper}, {Paegert},
  {De Lee}, {Torres}, {Latham}, {Muirhead}, {Dressing}, {Rojas-Ayala}, {Mann},
  {Fleming}, {Levine}, {Silvotti}, {Plavchan}, \& {the TESS Target Selection
  Working Group}}]{Stassun17}
{Stassun}, K.~G., {Oelkers}, R.~J., {Pepper}, J., {et~al.} 2017, ArXiv
  e-prints, arXiv:1706.00495

\bibitem[{{Str{\"u}der} {et~al.}(2001){Str{\"u}der}, {Briel}, {Dennerl},
  {Hartmann}, {Kendziorra}, {Meidinger}, {Pfeffermann}, {Reppin}, {Aschenbach},
  {Bornemann}, {Br{\"a}uninger}, {Burkert}, {Elender}, {Freyberg}, {Haberl},
  {Hartner}, {Heuschmann}, {Hippmann}, {Kastelic}, {Kemmer}, {Kettenring},
  {Kink}, {Krause}, {M{\"u}ller}, {Oppitz}, {Pietsch}, {Popp}, {Predehl},
  {Read}, {Stephan}, {St{\"o}tter}, {Tr{\"u}mper}, {Holl}, {Kemmer}, {Soltau},
  {St{\"o}tter}, {Weber}, {Weichert}, {von Zanthier}, {Carathanassis}, {Lutz},
  {Richter}, {Solc}, {B{\"o}ttcher}, {Kuster}, {Staubert}, {Abbey}, {Holland},
  {Turner}, {Balasini}, {Bignami}, {La Palombara}, {Villa}, {Buttler},
  {Gianini}, {Lain{\'e}}, {Lumb}, \& {Dhez}}]{Struder01}
{Str{\"u}der}, L., {Briel}, U., {Dennerl}, K., {et~al.} 2001, \aap, 365, L18

\bibitem[{{Su{\'a}rez-Andr{\'e}s} {et~al.}(2016){Su{\'a}rez-Andr{\'e}s},
  {Israelian}, {Gonz{\'a}lez Hern{\'a}ndez}, {Adibekyan}, {Delgado Mena},
  {Santos}, \& {Sousa}}]{SuarezAndres16}
{Su{\'a}rez-Andr{\'e}s}, L., {Israelian}, G., {Gonz{\'a}lez Hern{\'a}ndez},
  J.~I., {et~al.} 2016, \aap, 591, A69

\bibitem[{Takeda {et~al.}(2010)Takeda, Honda, Kawanomoto, Ando, \&
  Sakurai}]{Takeda10}
Takeda, Y., Honda, S., Kawanomoto, S., Ando, H., \& Sakurai, T. 2010, Astronomy
  and Astrophysics, 515, A93

\bibitem[{{Takeda} \& {Kawanomoto}(2005)}]{Takeda05Li}
{Takeda}, Y., \& {Kawanomoto}, S. 2005, \pasj, 57, 45

\bibitem[{Takeda {et~al.}(2007)Takeda, Taguchi, Yoshioka, Hashimoto, Aikawa, \&
  Kawanomoto}]{Takeda:2007p1531}
Takeda, Y., Taguchi, H., Yoshioka, K., {et~al.} 2007, PASJ, 59, 1127

\bibitem[{{Taylor}(2005)}]{Taylor05}
{Taylor}, M.~B. 2005, in Astronomical Society of the Pacific Conference Series,
  Vol. 347, Astronomical Data Analysis Software and Systems XIV, ed.
  P.~{Shopbell}, M.~{Britton}, \& R.~{Ebert}, 29

\bibitem[{Trevisan \& Barbuy(2014)}]{Trevisan14}
Trevisan, M., \& Barbuy, B. 2014, Astronomy {\&} Astrophysics, 570, A22

\bibitem[{Trevisan {et~al.}(2011)Trevisan, Barbuy, Eriksson, Gustafsson,
  Grenon, \& Pomp{\'e}ia}]{Trevisan:2011p6253}
Trevisan, M., Barbuy, B., Eriksson, K., {et~al.} 2011, A\&A, 535, 42

\bibitem[{{Tucci Maia} {et~al.}(2016){Tucci Maia}, {Ram{\'{\i}}rez},
  {Mel{\'e}ndez}, {Bedell}, {Bean}, \& {Asplund}}]{TucciMaia16}
{Tucci Maia}, M., {Ram{\'{\i}}rez}, I., {Mel{\'e}ndez}, J., {et~al.} 2016,
  \aap, 590, A32

\bibitem[{{Turnbull}(2015)}]{Turnbull15}
{Turnbull}, M.~C. 2015, ArXiv e-prints, arXiv:1510.01731

\bibitem[{{Unterborn} {et~al.}(2016){Unterborn}, {Dismukes}, \&
  {Panero}}]{Unterborn16}
{Unterborn}, C.~T., {Dismukes}, E.~E., \& {Panero}, W.~R. 2016, \apj, 819, 32

\bibitem[{Valenti \& Fischer(2005)}]{Valenti:2005p1491}
Valenti, J.~A., \& Fischer, D.~A. 2005, ApJS, 159, 141

\bibitem[{{Vaughan} \& {Preston}(1980)}]{Vaughan80}
{Vaughan}, A.~H., \& {Preston}, G.~W. 1980, \pasp, 92, 385

\bibitem[{{Voges} {et~al.}(1999){Voges}, {Aschenbach}, {Boller},
  {Br{\"a}uninger}, {Briel}, {Burkert}, {Dennerl}, {Englhauser}, {Gruber},
  {Haberl}, {Hartner}, {Hasinger}, {K{\"u}rster}, {Pfeffermann}, {Pietsch},
  {Predehl}, {Rosso}, {Schmitt}, {Tr{\"u}mper}, \& {Zimmermann}}]{Voges99}
{Voges}, W., {Aschenbach}, B., {Boller}, T., {et~al.} 1999, \aap, 349, 389

\bibitem[{{Voges} {et~al.}(2000){Voges}, {Aschenbach}, {Boller}, {Brauninger},
  {Briel}, {Burkert}, {Dennerl}, {Englhauser}, {Gruber}, {Haberl}, {Hartner},
  {Hasinger}, {Pfeffermann}, {Pietsch}, {Predehl}, {Schmitt}, {Trumper}, \&
  {Zimmermann}}]{Voges00}
---. 2000, \iaucirc, 7432

\bibitem[{{Vogt} {et~al.}(2015){Vogt}, {Burt}, {Meschiari}, {Butler}, {Henry},
  {Wang}, {Holden}, {Gapp}, {Hanson}, {Arriagada}, {Keiser}, {Teske}, \&
  {Laughlin}}]{Vogt15}
{Vogt}, S.~S., {Burt}, J., {Meschiari}, S., {et~al.} 2015, \apj, 814, 12

\bibitem[{{Wanke}(1981)}]{Wanke81}
{Wanke}, H. 1981, Philosophical Transactions of the Royal Society of London
  Series A, 303, 287

\bibitem[{{Weisskopf} {et~al.}(2000){Weisskopf}, {Tananbaum}, {Van Speybroeck},
  \& {O'Dell}}]{Weisskopf00}
{Weisskopf}, M.~C., {Tananbaum}, H.~D., {Van Speybroeck}, L.~P., \& {O'Dell},
  S.~L. 2000, in \procspie, Vol. 4012, X-Ray Optics, Instruments, and Missions
  III, ed. J.~E. {Truemper} \& B.~{Aschenbach}, 2--16

\bibitem[{{Wielen}(1974)}]{Wielen74}
{Wielen}, R. 1974, Highlights of Astronomy, 3, 395

\bibitem[{{Wilson}(1978)}]{Wilson78}
{Wilson}, O.~C. 1978, \apj, 226, 379

\bibitem[{{Wright} {et~al.}(2004){Wright}, {Marcy}, {Butler}, \&
  {Vogt}}]{Wright04}
{Wright}, J.~T., {Marcy}, G.~W., {Butler}, R.~P., \& {Vogt}, S.~S. 2004, \apjs,
  152, 261

\bibitem[{{Wright} {et~al.}(2011){Wright}, {Drake}, {Mamajek}, \&
  {Henry}}]{Wright11}
{Wright}, N.~J., {Drake}, J.~J., {Mamajek}, E.~E., \& {Henry}, G.~W. 2011,
  \apj, 743, 48

\bibitem[{Yan {et~al.}(2016)Yan, Shi, Nissen, \& Zhao}]{Yan16}
Yan, H.~L., Shi, J.~R., Nissen, P.~E., \& Zhao, G. 2016, Astronomy {\&}
  Astrophysics, 585, A102

\bibitem[{{\v{Z}}enovien{\.{e}} {et~al.}(2015){\v{Z}}enovien{\.{e}},
  Tautvai{\v{s}}ien{\.{e}}, Nordstr̦m, Stonkut{\.{e}}, \&
  Barisevi{\v{c}}ius}]{Zenoviene15}
{\v{Z}}enovien{\.{e}}, R., Tautvai{\v{s}}ien{\.{e}}, G., Nordstr̦m, B.,
  Stonkut{\.{e}}, E., \& Barisevi{\v{c}}ius, G. 2015, Astronomy {\&}
  Astrophysics, 576, A113

\bibitem[{{Zhao} {et~al.}(2016){Zhao}, {Mashonkina}, {Yan}, {Alexeeva},
  {Kobayashi}, {Pakhomov}, {Shi}, {Sitnova}, {Tan}, {Zhang}, {Zhang}, {Zhou},
  {Bolte}, {Chen}, {Li}, {Liu}, \& {Zhai}}]{Zhao16}
{Zhao}, G., {Mashonkina}, L., {Yan}, H.~L., {et~al.} 2016, \apj, 833, 225

\end{thebibliography}
\end{document}